\documentclass[11pt,a4paper]{article}
\pdfoutput=1
\RequirePackage{snapshot}  % required by bundledoc
\usepackage{jheppub}
\usepackage{color}
\usepackage{graphicx}
\usepackage{xspace}
\usepackage{booktabs}
\usepackage[caption=false]{subfig}

\makeatletter
\g@addto@macro\bfseries{\boldmath}
\makeatother

\newcommand{\zcut}{z_\text{cut}}

\newcommand{\nPU}{n_\text{PU}}

\newcommand{\RSDinf}{\text{RSD}_\infty}
\newcommand{\RSDN}{\text{RSD}_N}

\newcommand{\RSDNarg}{\text{RSD}_N(\beta,\zcut)}

\newcommand{\GeV}{\,\mathrm{GeV}}

\newcommand{\pythia}[1]{\textsc{Pythia\xspace #1}}

\newcommand{\fastjet}[1]{\textsc{FastJet\xspace #1}}

\newcommand{\as}{\alpha_s}

\definecolor{darkgreen}{rgb}{0,0.5,0}
\definecolor{darkblue}{rgb}{0,0,0.7}
\definecolor{darkred}{rgb}{0.5,0,0.0}
\definecolor{darkorange}{rgb}{0.8,0.4,0.0}
\definecolor{blue}{rgb}{0.0,0.0,1.0}

\DeclareRobustCommand{\Sec}[1]{Sec.~\ref{#1}}
\DeclareRobustCommand{\Secs}[2]{Secs.~\ref{#1} and \ref{#2}}

\DeclareRobustCommand{\App}[1]{App.~\ref{#1}}

\DeclareRobustCommand{\Fig}[1]{Fig.~\ref{#1}}

\DeclareRobustCommand{\Eq}[1]{Eq.~(\ref{#1})}

\DeclareRobustCommand{\Ref}[1]{Ref.~\cite{#1}}
\DeclareRobustCommand{\Refs}[1]{Refs.~\cite{#1}}
\preprint{MIT--CTP/4988}

\title{Recursive Soft Drop}

\author[1]{Fr\'ed\'eric A. Dreyer,}
\author[2]{Lina Necib,}
\author[3]{Gregory Soyez,}
\author[1]{and Jesse Thaler}

\affiliation[1]{Center for Theoretical Physics,
  Massachusetts Institute of Technology,
  Cambridge, MA 02139, USA}
\affiliation[2]{Walter Burke Institute for Theoretical Physics,
California Institute of Technology, Pasadena, CA 91125, USA}
\affiliation[3]{IPhT, CEA Saclay,
  CNRS UMR 3681,
  F-91191 Gif-sur-Yvette cedex, France}

\emailAdd{fdreyer@mit.edu}
\emailAdd{lnecib@caltech.edu}
\emailAdd{gregory.soyez@ipht.fr}
\emailAdd{jthaler@mit.edu}

\keywords{QCD, Hadronic Colliders, Standard Model, Jets}

\abstract{%
  We introduce a new jet substructure technique called Recursive Soft
  Drop, which generalizes the Soft Drop algorithm to have multiple
  grooming layers.
  Like the original Soft Drop method, this new recursive variant
  traverses a jet clustering tree to remove soft wide-angle
  contamination.
  By enforcing the Soft Drop condition $N$ times, Recursive Soft Drop
  improves the jet mass resolution for boosted hadronic objects like
  $W$ bosons, top quarks, and Higgs bosons.
  We further show that this improvement in mass resolution persists
  when including the effects of pileup, up to large pileup
  multiplicities.
  In the limit that $N$ goes to infinity, the resulting groomed jets
  formally have zero catchment area.
  As an alternative approach, we present a bottom-up version of
  Recursive Soft Drop which, in its local form, is similar to
  Recursive Soft Drop and which, in its global form, can be used to
  perform event-wide grooming.  
}

\begin{document}
\maketitle

%======================================================================
\section{Introduction}

As the Large Hadron Collider (LHC) collides protons at the highest energies
accessible in a laboratory setting, electroweak-scale resonances are
routinely produced with transverse momenta far exceeding their rest
mass.
These highly boosted objects will generate collimated hadronic decays,
which are often reconstructed as a single fat
jet.
Due to the differences in their radiation patterns, fat jets
originating from boosted objects can be distinguished from ordinary
quark and gluon jets by studying their substructure.
Since the start of the experimental program of the LHC, jet
substructure has matured into a highly active field of
research~\cite{Abdesselam:2010pt,Altheimer:2012mn,Altheimer:2013yza,Adams:2015hiv,Cacciari:2015jwa,Larkoski:2017jix,LH2017}.
First introduced in the pioneering studies of
\Refs{Seymour:1991cb,Seymour:1993mx,Butterworth:2002tt,Butterworth:2007ke}, jet
substructure was revived by seminal work showing its potential
application in the search for a light Higgs boson decaying to bottom
quarks~\cite{Butterworth:2008iy}.

By now, a variety of tools use jet substructure to tag boosted objects
and mitigate contamination from poorly modeled contributions such as
underlying event and pileup
\cite{Thaler:2008ju,Kaplan:2008ie,Krohn:2009th,Ellis:2009su,Ellis:2009me,Plehn:2009rk,Kim:2010uj,Thaler:2010tr,Thaler:2011gf,Larkoski:2013eya,Chien:2013kca,Larkoski:2014gra,Moult:2016cvt,Feige:2012vc,Field:2012rw,Dasgupta:2013ihk,Dasgupta:2013via,Larkoski:2014pca,Dasgupta:2015yua,Seymour:1997kj,Li:2011hy,Larkoski:2012eh,Jankowiak:2012na,Chien:2014nsa,Chien:2014zna,Isaacson:2015fra,Krohn:2012fg,Waalewijn:2012sv,Larkoski:2014tva,Larkoski:2014wba,Procura:2014cba,Bertolini:2015pka,Bhattacherjee:2015psa,Larkoski:2015kga,Dasgupta:2015lxh,Dasgupta:2016ktv,Salam:2016yht,Frye:2016okc,Frye:2016aiz,Kang:2016ehg,Hornig:2016ahz,Marzani:2017mva,Marzani:2017kqd,Chien:2017xrb,Chien:2018dfn}, which have already
found numerous experimental
applications~\cite{Chatrchyan:2013vbb,Aad:2013gja,Khachatryan:2014vla,Chatrchyan:2012sn,CMS:2013cda,Aad:2015cua,Aad:2015lxa,ATLAS-CONF-2015-035,Aad:2015rpa,Aad:2015hna,ATLAS-CONF-2016-002,ATLAS-CONF-2016-039,ATLAS-CONF-2016-034,CMS-PAS-TOP-16-013,CMS-PAS-HIG-16-004,CMS:2011bqa,Fleischmann:2013woa,Pilot:2013bla,TheATLAScollaboration:2013qia,Chatrchyan:2012ku,CMS-PAS-B2G-14-001,CMS-PAS-B2G-14-002,Khachatryan:2015axa,Khachatryan:2015bma,Aad:2015owa,Aaboud:2016okv,Aaboud:2016trl,Aaboud:2016qgg,ATLAS-CONF-2016-055,ATLAS-CONF-2015-071,ATLAS-CONF-2015-068,CMS-PAS-EXO-16-037,CMS-PAS-EXO-16-040,Khachatryan:2016mdm,CMS-PAS-HIG-16-016,CMS-PAS-B2G-15-003,CMS-PAS-EXO-16-017}.
One particular technique that has emerged both as a powerful
substructure probe and as an analytically tractable approach is the
modified Mass Drop Tagger (mMDT)~\cite{Dasgupta:2013ihk}, and its later
extension, Soft Drop (SD)~\cite{Larkoski:2014wba}.
The SD procedure takes an initial jet with radius $R_0$,
reclusters its constituents with the Cambridge/Aachen (C/A) algorithm
\cite{Dokshitzer:1997in,Wobisch:1998wt}, and removes soft wide-angle
emissions that do not satisfy the SD condition, defined as
\begin{equation}
  \label{eq:SD-crit-intro}
  \frac{\min(p_{t,1}, p_{t,2})}{p_{t,1} + p_{t,2}}
  > \zcut \left(\frac{\Delta R_{12}}{R_0}\right)^\beta\,,
\end{equation}
where the notation will be reviewed below.  This method has been used
in a variety of analyses at the LHC, including jet mass and transverse
momentum measurements in dijet
events~\cite{CMS-PAS-SMP-16-010,Aaboud:2017qwh}, vector resonance and
dark matter
searches~\cite{Sirunyan:2017dnz,Aaboud:2017eta,Sirunyan:2018ivv,Sirunyan:2018gka},
and boosted $H\rightarrow b\bar{b}$ searches~\cite{Sirunyan:2017dgc}.
It has also been used as a powerful probe of the QCD splitting
function, both in proton-proton
collision~\cite{Larkoski:2017bvj,Tripathee:2017ybi} and in heavy ion
collisions~\cite{Sirunyan:2017bsd,Caffarri:2017bmh,Kauder:2017mhg},
where the shared momentum fraction $z_g$ provides a handle on medium
effects~\cite{Chien:2016led,Milhano:2017nzm,Mehtar-Tani:2016aco}.
Because grooming with mMDT/SD removes complications due to
unassociated wide-angle emissions, it has also allowed analytic
calculations of the groomed jet mass to reach unprecedented
accuracies~\cite{Frye:2016okc,Frye:2016aiz,Marzani:2017mva,Marzani:2017kqd}.

In this paper, we introduce a recursive extension of the SD
algorithm---aptly named Recursive Soft Drop (RSD)---where SD is
reapplied along the C/A clustering history until a specified number
$N$ of SD conditions have been satisfied.
We focus on jet grooming with RSD, taking an angular
exponent $\beta \ge 0$.
The case $N=0$ involves no jet grooming, the case $N=1$ corresponds to
the original SD procedure, and the cases $N \ge 2$ are well-suited to
multi-prong boosted objects.
Like the original SD, RSD is stable under hadronization and underlying
event effects, but RSD is able to provide improved jet mass resolution
for signals such as boosted 2-prong $W$ bosons, 3-prong top quarks,
and 4-prong Higgs bosons.
Intriguingly, in the $N \to \infty$ limit, groomed jets from RSD formally have
zero catchment area~\cite{Cacciari:2008gn}, 
a feature that suggests that RSD would be well suited for pileup
mitigation.

\begin{figure}[t]
  \centering
  \subfloat[]{\includegraphics[width=0.45\textwidth]{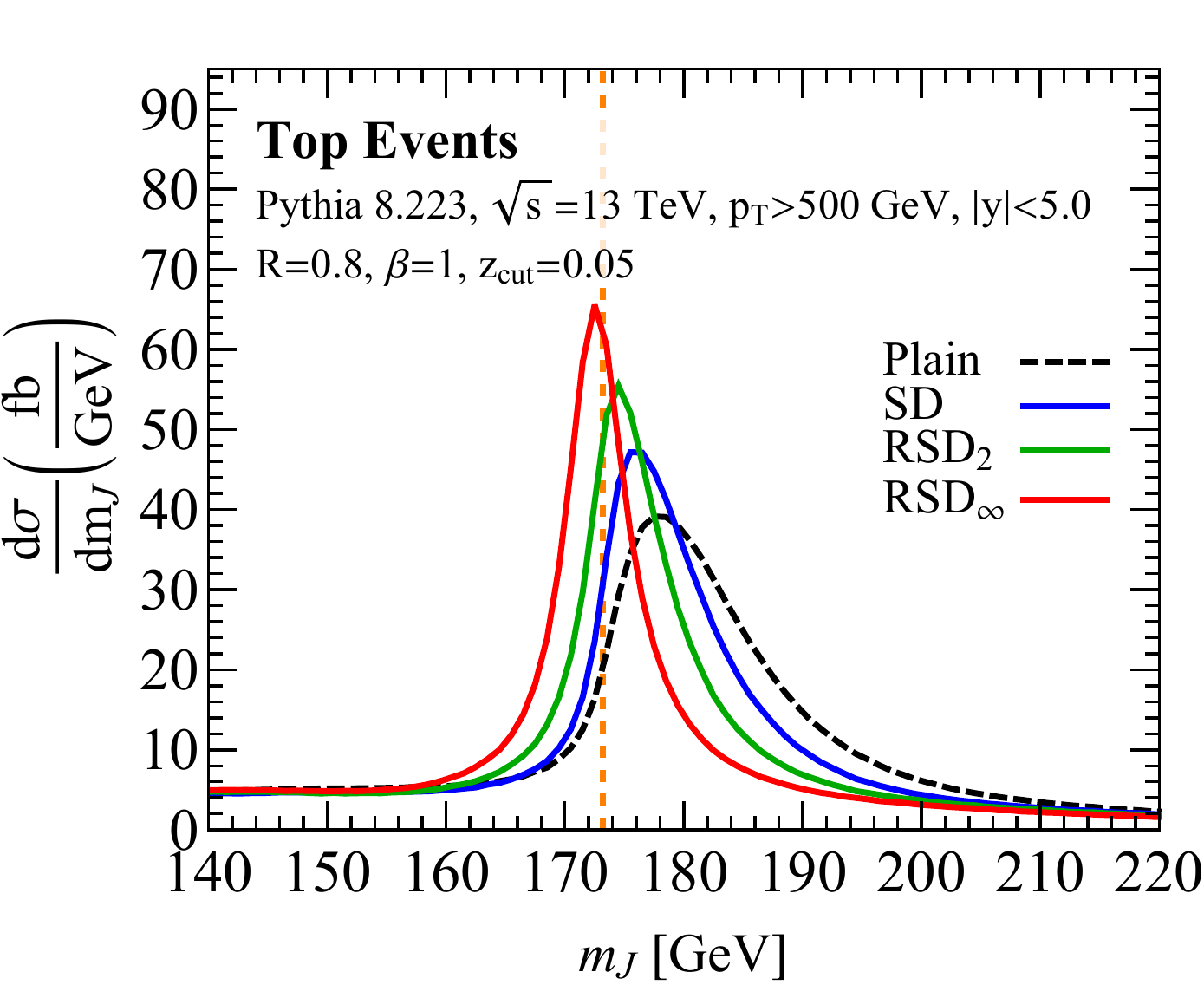}%
    \label{fig:summary_top}}%
  \qquad
  \subfloat[]{\includegraphics[trim={0cm 0cm -0cm 0cm},clip,width=0.45\textwidth]{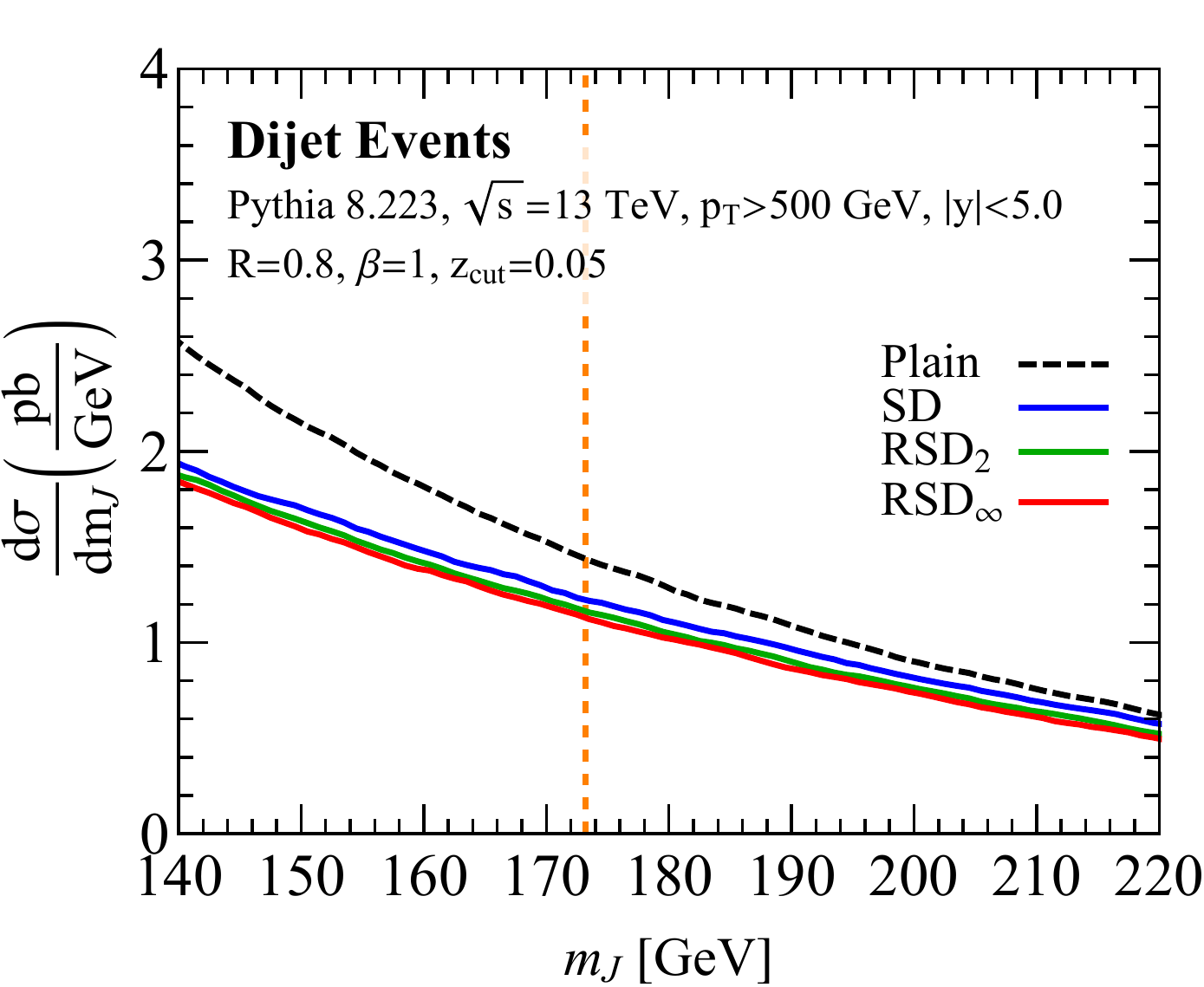}%
    \label{fig:summary_dijet}}%
  \caption{Behavior of RSD in (a) boosted top signals and (b)
    corresponding dijet backgrounds, using \pythia 8.223 and $R=0.8$
    anti-$k_t$ jets.  The benchmark SD condition in
    \Eq{eq:SD-crit-intro} uses $\beta = 1$ and $\zcut = 0.05$.}
  \label{fig:summary}
\end{figure}

We focus our attention here on the phenomenological applications of
RSD, leaving a detailed study of its analytical properties to future
work.
The behavior of RSD is summarized in \Fig{fig:summary} in the case of distinguishing 
boosted top quark signals from quark/gluon jet
backgrounds.
As the number of RSD layers increases, the top mass peak is better
resolved, with the best performance (for this choice of SD condition)
achieved in the infinite $N$ limit, hereafter labeled RSD$_\infty$.
For quark/gluon jets, RSD has a much smaller impact on the groomed
mass, but there are still substantial gains in top tagging performance
just from the increased signal mass resolution.
In general, any application at the LHC suitable for SD is also
suitable for RSD, with the possibility of loosening the SD condition
while increasing the number of layers $N$ to balance performance and
robustness.
As a concrete example, we show how to use RSD$_\infty$ in combination
with either the SoftKiller~\cite{Cacciari:2014gra} or the area--median
method~\cite{Cacciari:2007fd,Cacciari:2008gn} to mitigate pileup in
high-luminosity conditions.

The rest of this paper is organized as follows.
We introduce the RSD algorithm in~\Sec{sec:recSD-alg}, and describe
its basic features in~\Sec{sec:basic}, such as its robustness to
non-perturbative effects and the $N\rightarrow\infty$ limit of
zero-area jets.
In~\Sec{sec:mass-resol}, we show how RSD improves jet mass resolution
for boosted resonances with hadronic decays, and we present a brief
case study of boosted top tagging.
In~\Sec{sec:pileup-mitigation}, we discuss the application of RSD to
pileup mitigation.
We present an alternative version of the RSD algorithm in
\Sec{sec:BUSD}, where a bottom-up strategy can be
applied locally (at the jet level) or globally (at the event level).
We conclude in \Sec{sec:conclusions}, leaving additional studies to
the appendices.

%======================================================================
\section{The Recursive Soft Drop algorithm}
\label{sec:recSD-alg}

%======================================================================

Since RSD is a generalization of SD, we first summarize the SD
algorithm~\cite{Larkoski:2014wba} in \Sec{sec:SD}, and introduce the
multi-layer RSD$_N$ algorithm in \Sec{sec:recursive-SD}.  
We present a more aggressive RSD variant in \Sec{sec:dynamicR}.
Like all jet
grooming procedures, RSD$_N$ removes wide-angle soft radiation within
a jet, with the new feature that the meaning of ``wide angle'' is
determined recursively.

%======================================================================
\subsection{Review of Soft Drop} 
\label{sec:SD}
%======================================================================

The original SD algorithm starts from any jet, where the constituents are reclustered into a C/A angular-ordered tree~\cite{Dokshitzer:1997in,Wobisch:1998wt}.
The degree of jet grooming depends on two parameters: the minimum energy fraction
$z_{\text{cut}}$ allowed for the softer branch, and an angular exponent
$\beta$ defining how much collinear radiation is removed by the
grooming procedure.
It is also convenient to introduce a reference angular scale $R_0$ (absorbable into the definition of $z_{\text{cut}}$), which is typically set to the initial jet clustering radius $R$.
We denote by $p_{t,i}$ the transverse momentum of the $i$-th subjet, and
by $\Delta R_{ij}$ the rapidity-azimuth distance between the $i$-th and $j$-th
subjets.

The SD algorithm proceeds as follows:
\begin{enumerate}
\item Undo the last C/A clustering step of the jet $j$ and label the
  two parent subjets as $j_1$ and $j_2$.
\item If these subjets pass the SD condition,
  \begin{equation}
    \label{eq:SD-crit}
    z_{12} > \zcut \left(\frac{\Delta R_{12}}{R_0}\right)^\beta\,, \qquad z_{12} \equiv \frac{\min(p_{t,1},p_{t,2})}{p_{t,1} + p_{t,2}},
  \end{equation}
  then the procedure stops and the SD jet $j$ is returned.
\item Otherwise, the softer subjet (by $p_t$) is removed and the algorithm
  iterates on the new jet $j$ defined by the harder subjet.
\item If $j$ has no further subjets, either terminate without
  returning a jet (tagging mode) or define $j$ to be the SD jet
  (grooming mode).
\end{enumerate}
As explained in \Ref{Larkoski:2014wba}, this algorithm is infrared and collinear (IRC) safe for $\beta>0$ in grooming mode, though it remains
Sudakov safe \cite{Larkoski:2013paa,Larkoski:2015lea} for $\beta\rightarrow 0$.\footnote{In tagging mode, SD is IRC safe for $\beta < 0$.  If a non-trivial mass cut is applied, SD is also IRC safe in tagging mode for $\beta  = 0$.}
The limits $\zcut\rightarrow 0$ or $\beta\rightarrow \infty$ return an
ungroomed jet.
Finally, the limit $\beta\to 0$ corresponds to mMDT~\cite{Dasgupta:2013ihk}.

%======================================================================
\subsection{Introducing Recursive Soft Drop}
\label{sec:recursive-SD}
%======================================================================

\begin{figure}[t]
  \centering
    \includegraphics[trim={12cm 10cm 6cm 6cm},clip,width=0.6\textwidth]{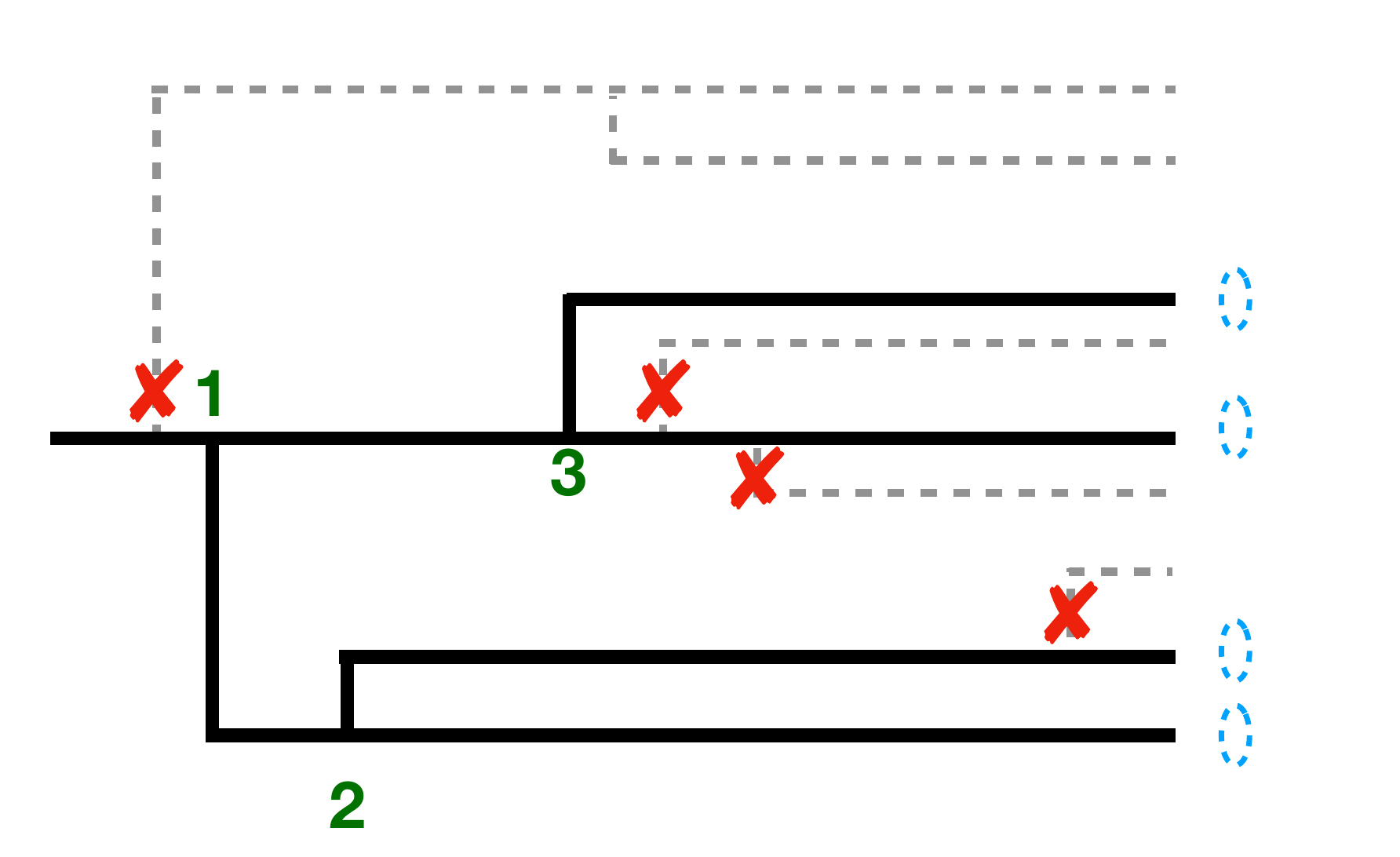}%
  \caption{ Schematic depiction of the RSD$_N$ algorithm, where a C/A clustering tree is declustered until \Eq{eq:SD-crit} is satisfied $N$ times.  For $N \le 3$, the grooming stops at the numbered branch $N$, such that $N = 1$ corresponds to the original SD algorithm.  The grooming of all dashed gray lines is achieved for $N>3$.}
  \label{fig:RSD}
\end{figure}

As depicted in \Fig{fig:RSD}, RSD$_{N}$ grooms a jet by applying $N$ layers of SD declustering,
iterating through the full jet clustering tree.
This is achieved by ordering all branches by the $\Delta R_{ij}$
separation of their constituents, and iterating through the tree structure
by taking the branch with the most widely-separated constituents at
each step.

More explicitly, starting from a C/A-reclustered jet:
\begin{enumerate}
\item Set the list of branches to a single element: the initial jet.
\item Take the remaining branch whose two parent subjets have the
  widest separation in $\Delta R$, and label these $j_1$ and $j_2$.%
  \footnote{During the first iteration, this step is of course trivial,
    since there is only one branch to the C/A tree.}
  Remove that branch from the list of branches.
\item If the two subjets pass the SD condition in~\Eq{eq:SD-crit},
  keep both subjets as new branches; otherwise, remove the softer of
  the two subjets and keep the hardest as a new branch.  
\item Iterate this process until the SD
  condition~in \Eq{eq:SD-crit} has been met $N$ times, or until the
   C/A tree has been fully recursed through (using the same
   definitions of grooming/tagging mode as in ordinary SD).
   The groomed jet is then made of all the remaining branches.
\end{enumerate}
After $N$ iterations where the SD condition is satisfied, one
obtains a groomed jet constructed of $(N+1)$ subjets.%
\footnote{Alternatively, one could consider a fixed-depth recursion
  instead, where the SD condition is applied $N$ times on each branch
  of the clustering tree, resulting in (up to) $2^N$ prongs. This coincides with the variable depth algorithm in the $N\rightarrow \infty$ limit, but will differ at finite $N$ due to the removal of small angle emissions on the subleading branch.}
For $N=0$, this procedure returns the initial ungroomed jet, while
for $N=1$ it is equivalent to the original SD algorithm.
We use $\RSDNarg$, or simply $\RSDN$, to denote
RSD grooming with $N$ iterations and parameters $\beta$ and $\zcut$, such
that $\text{RSD}_0 = \mathbf{1}$ and $\text{RSD}_1 = \text{SD}$.
For fully recursive SD grooming with $N=\infty$, we use RSD$_\infty$.%
\footnote{Note that RSD$_\infty$ is only well defined in grooming
  mode, and, therefore, only IRC safe for $\beta>0$.}

In this paper, we only study RSD with $\beta \ge 0$ in grooming mode.
In principle, one could also use $\RSDN(\beta<0,\zcut)$ to define an
$(N+1)$-prong tagger.
For example, one could use RSD$_2$ with $\beta<0$ as a top tagger,
much in the way $\text{SD} = \text{RSD}_1$ can be used for boosted $W$
tagging (see Section~7 of \Ref{Larkoski:2014wba}). 
Ultimately, though, we find that RSD$_\infty$ (with $\beta \ge 0$ in grooming mode) shows good overall tagging performance, making it our recommended default algorithm.

It is worth noting that RSD bears some resemblance to the Iterated Soft Drop (ISD) procedure
recently introduced in \Ref{Frye:2017yrw} for quark/gluon
discrimination.
A key difference, however, is that RSD follows both branches of the
clustering tree, while ISD limits itself to traversing only the harder
branch.
Both RSD and ISD, along with mMDT and SD, are implemented in
{\tt{RecursiveTools}} ($\ge$2.0.0) included as part of
\texttt{fastjet-contrib}~\cite{fjcontrib}.

%======================================================================
\subsection{Dynamic $R_0$ for aggressive grooming}
\label{sec:dynamicR}

In the default RSD algorithm, the $R_0$ in \Eq{eq:SD-crit} is fixed to
the initial jet radius (henceforth denoted {\it fixed $R_0$} mode).
One can instead take a more aggressive approach, in which one updates
$R_0$ dynamically during the grooming procedure.

In this {\it dynamic} $R_0$ mode, at each step of the process
where two particles $i,j$ meet the SD condition, the $R_0$ value is
updated to $R_0=\Delta R_{ij}$, with $R_0$ being kept independent on
each branch of the C/A clustering tree.
By decreasing $R_0$ in each grooming step for $\beta > 0$, one imposes a more
stringent requirement on the momentum fraction.
This yields a more aggressive grooming strategy for all $N>1$.

For most practical purposes, the dynamic $R_0$ variant behaves quite
similarly to RSD with a smaller $\beta$ value.
As we will see in \Sec{sec:PU-area}, though, it does have some
specific advantages for pileup mitigation with area--median
subtraction.

\section{Basic properties of Recursive Soft Drop}
\label{sec:basic}

We perform a variety of parton shower studies in this paper to
highlight the features of RSD.
In this section, as well as in the more detailed studies in
\Secs{sec:mass-resol}{sec:pileup-mitigation}, we always generate
$\sqrt{s} = 13$~TeV proton-proton collisions in \pythia~8.223
\cite{Sjostrand:2006za,Sjostrand:2007gs,Sjostrand:2014zea} with the
default 4C tune.
To reduce computation time, we turn off $\pi^0$ and $B$-hadrons decays (still letting
other hadrons decay).\footnote{One might wonder whether $\pi^0$ and $B$-hadrons decays
  would affect our SoftKiller pileup study in \Sec{sec:mass_PU}. 
  Since these decays increase the particle multiplicity, a slightly smaller
  value of the $a_{\text{SK}}$ would be preferable (see e.g. Figs.~12.3 and
  13.1 of \Ref{Soyez:2018opl}), but the qualitative features would remain the same.}
Jet clustering is performed with \fastjet{3.2.1} \cite{Cacciari:2011ma},
using the anti-$k_t$ algorithm \cite{Cacciari:2008gp} with the
default $E$-scheme recombination and a jet radius $R=0.8$.
We then select jets that have transverse momentum $p_T > 500~\GeV$ and
rapidity $|y|<5$.

To demonstrate the key similarities and differences between SD and RSD$_N$, we discuss the groomed radii and splitting scales in \Sec{sec:radius}, the robustness to non-perturbative effects in
\Sec{sec:NP-effects}, and the $N\rightarrow \infty$ limit of zero-area jets in \Sec{sec:zero-area}.
In \App{sec:fixed_order}, we present fixed-order studies of the RSD jet mass distribution to order $\alpha_s$ and $\alpha_s^2$.

%======================================================================
\subsection{Groomed radii and momentum fractions}
\label{sec:radius}
%======================================================================

In addition to grooming a jet, RSD defines a range of new jet
observables.
At each SD step $i$, one can define the groomed radius $R_{g,i}$
and momentum fraction $z_{g,i}$, equal to the $\Delta R_{12}$ and
$z_{12}$ values for the corresponding branch that passes
\Eq{eq:SD-crit}.
For $\beta > 0$, the $R_{g,i}$ observables are IRC safe, while the $z_{g,i}$ are in general Sudakov safe \cite{Larkoski:2015lea}.
Thus, after $N$ layers of RSD grooming, we obtain $N$ pairs of $\{R_{g,i}, z_{g,i} \}$ values containing information about the grooming history of the jet.
The values obtained for $i = 1$ are identical to the ones obtained from ordinary SD.%
\footnote{The ISD procedure in \Ref{Frye:2017yrw} also returns a set of $\{R_{g,i}, z_{g,i} \}$ pairs, but they are in general different from RSD, even for the same $\beta$ and $\zcut$ values, since ISD only follows the trunk of the clustering tree.}  

\begin{figure}[t]
  \centering
  \subfloat[]{%
    \includegraphics[width=0.45\textwidth]{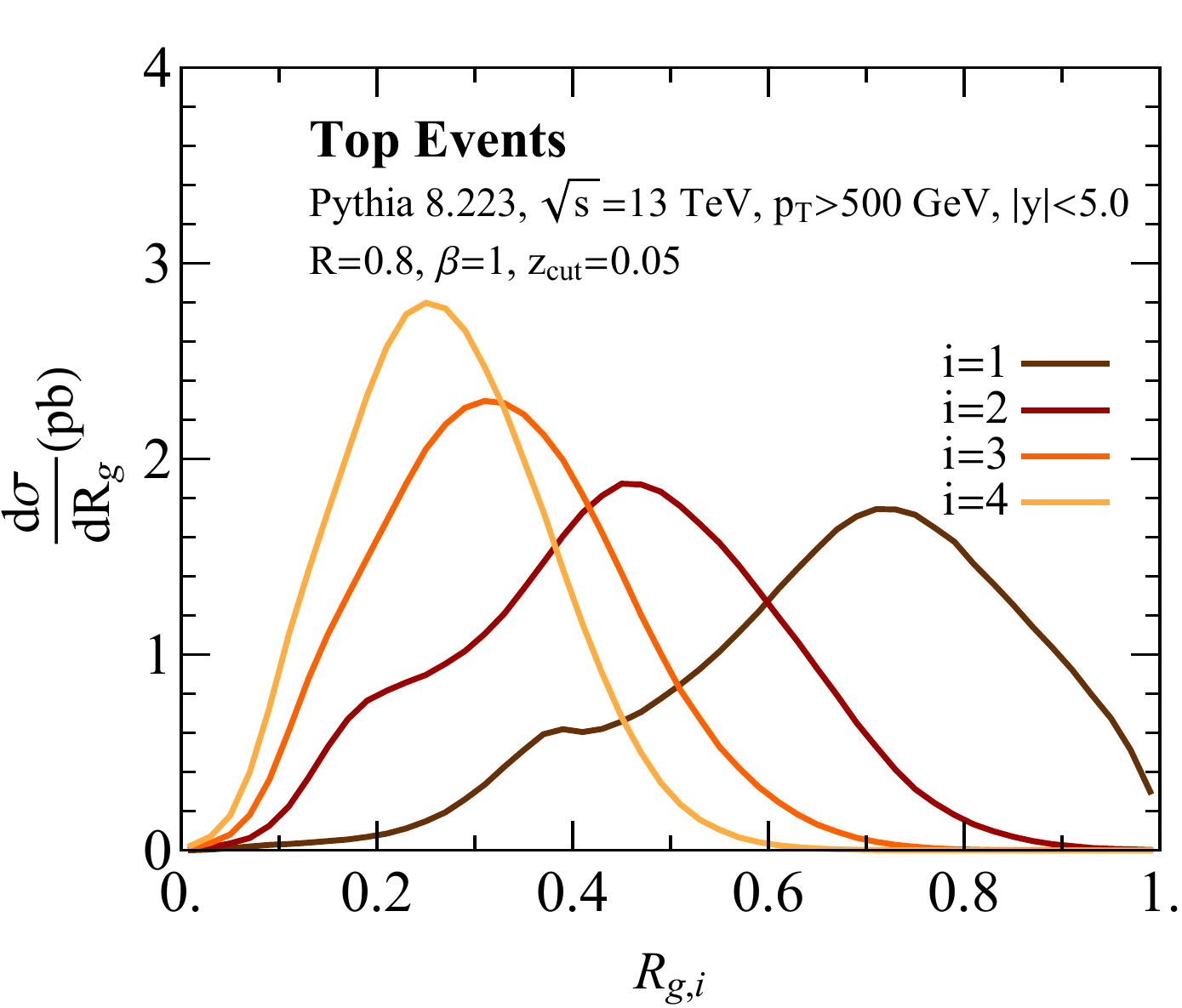}%
    \label{fig:top-rsd-internal_Rg}}% 
  \qquad
  \subfloat[]{%
    \includegraphics[width=0.45\textwidth]{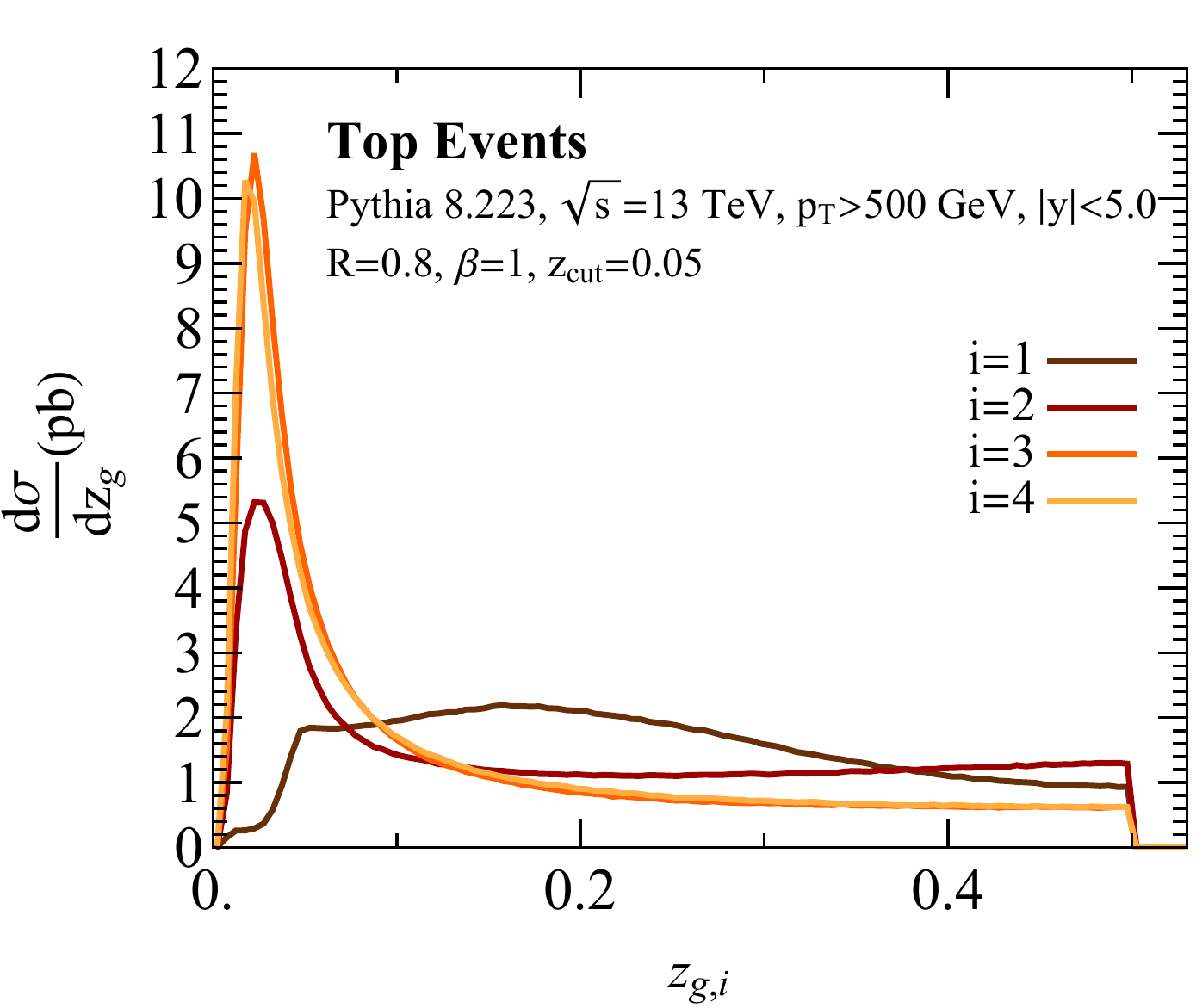}%
    \label{fig:top-rsd-internal_zg}}%
  \caption{Distributions in the top jet sample of (a) groomed radius and (b) groomed momentum fraction for $i=1,2,3,4$.}
  \label{fig:top-rsd-internal}
\end{figure}

We now consider boosted top quark events from the process $pp\rightarrow t\bar{t}$, forcing the tops to decay in the hadronic channel.
In \Fig{fig:top-rsd-internal_Rg}, we show the $R_{g,i}$ distributions from RSD$_i$ with $i \in \{1,2,3,4\}$, taking $\beta = 1$ and  $\zcut = 0.05$ as baseline parameters. 
As expected, the groomed radius of the jet decreases as more layers of RSD are applied.
There is a small kink structure at lower values of $R_g$ for $i=\{1,2\}$ due to the presence of
$W$ jets in the top quark decay.
In \Fig{fig:top-rsd-internal_zg} we show the $z_{g,i}$ distributions, and find that the momentum
fraction of the groomed jet is also peaked at values closer to zero as $i$ increases.
The sharp cut at 0.5 is because $z_g$ is defined as the relative momentum fraction of the softer subjet, which can be at most half.
For $i=1$, the $z_{g,i}$ distribution has a nontrivial structure from cases where the clustering history differs from the expected $t \to b W$ topology.

The case of $N = 2$ yields a three-pronged grooming strategy, which may be useful for the study of boosted top
decays, and we explore this possibility in \Sec{sec:boosted-toptag}.
Alternatively, one can use the $\{R_{g,i}, z_{g,i} \}$ values directly to discriminate
signal events from backgrounds.
For example, we can use ratios like $R_{g,3}/R_{g,2}$ as a probe for boosted top jets, somewhat analogous to the use of the
$N$-subjettiness ratio $\tau_{32}$~\cite{Thaler:2010tr,Thaler:2011gf}.
Similarly, one can use the $z_{g,1}$, and $z_{g,2}$ observables to distinguish QCD-like $1\rightarrow 2$ parton splittings (see e.g.~\cite{Larkoski:2017bvj}) from hard $t \to bW$ and $W \to q \overline{q}'$ decays.
In practice, though, top taggers built from $R_{g,i}$ and $z_{g,i}$ do
not seem to perform quite as well as $N$-subjettiness (with
RSD${}_\infty$ grooming), but might remain useful inputs for
multivariate analyses, depending on how much $R_{g,i}$ and $z_{g,i}$ are correlated with $N$-subjettiness.

%======================================================================
\subsection{Non-perturbative effects}
\label{sec:NP-effects}
%======================================================================

\begin{figure}[t]
  \centering
  \subfloat[]{%
    \includegraphics[width=0.45\textwidth]{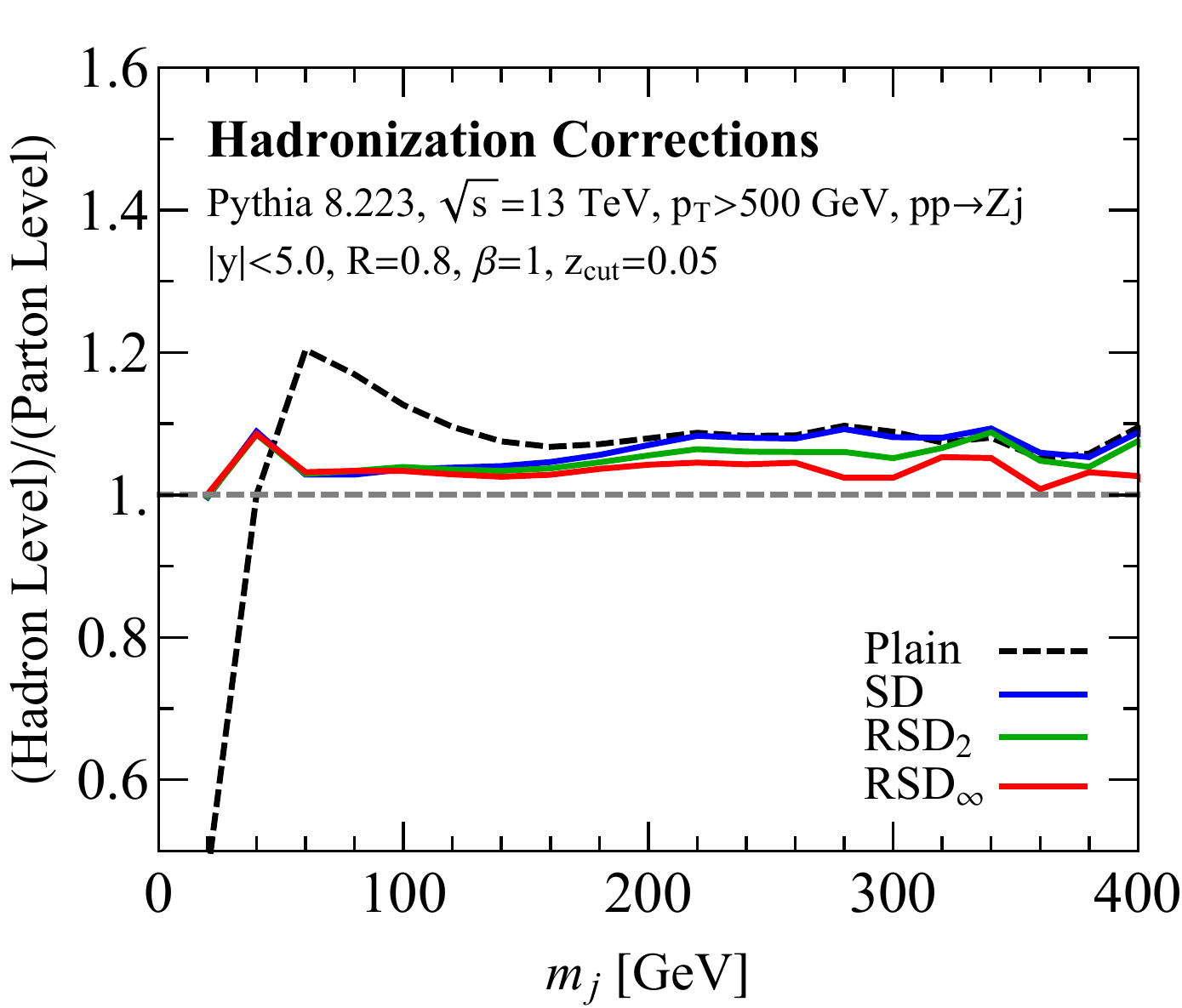}%
    \label{fig:m_NP_hadr}}%
  \qquad
  \subfloat[]{%
    \includegraphics[width=0.45\textwidth]{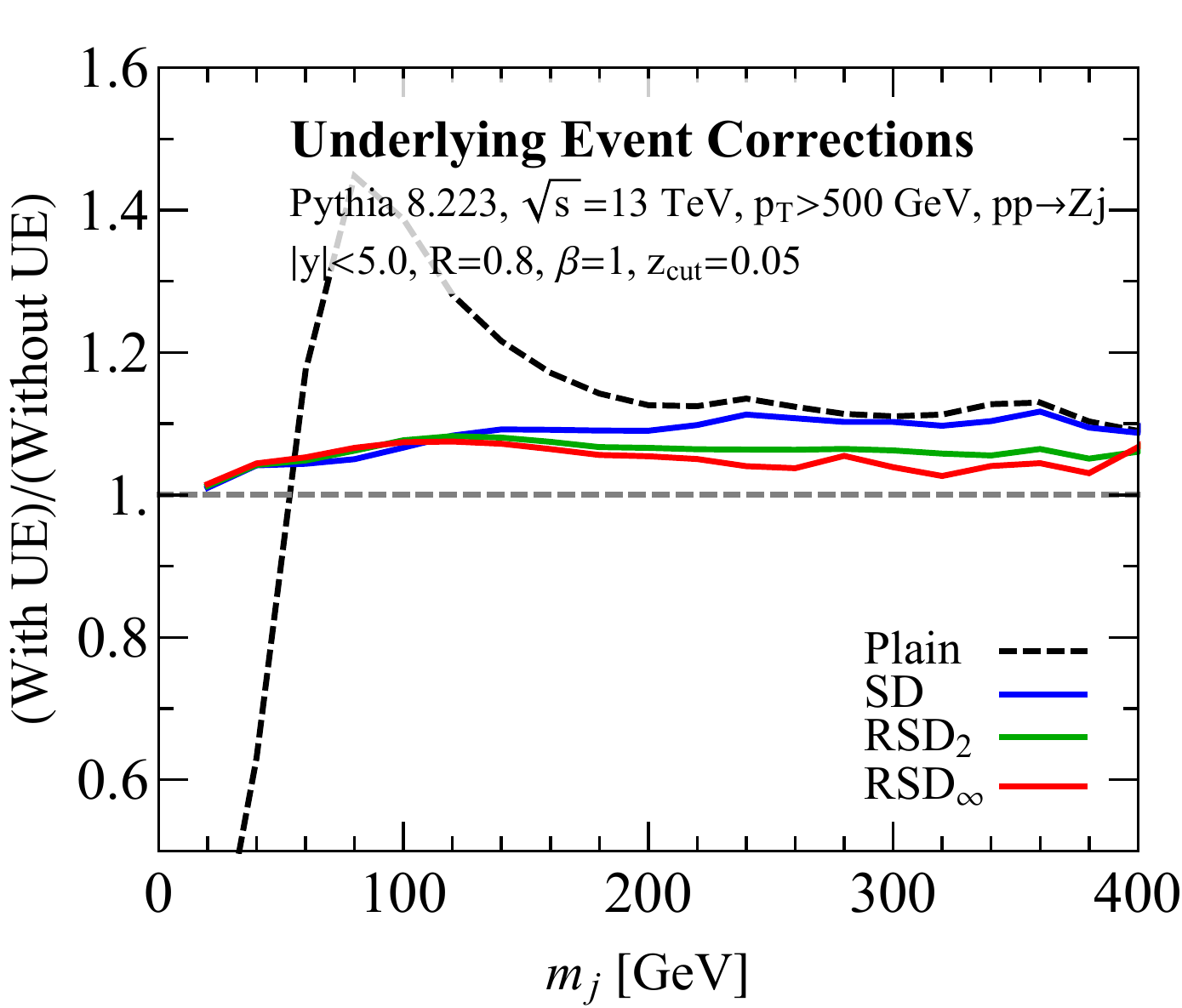}%
    \label{fig:m_NP_UE}}
  \caption{Study of (a) hadronization and (b) underlying event corrections for different $N$ values, with $\beta = 1$ and $\zcut = 0.05$.  Shown are ratios of the groomed jet mass spectrum as non-perturbative effects are included in \textsc{Pythia}.}
  \label{fig:m_NP}
\end{figure}

Analytical control in jet substructure is mainly limited to perturbative QCD effects.
Because internal jet properties probe very exclusive
kinematic regions, however, it is not uncommon for non-perturbative effects to
yield substantial corrections to perturbative predictions.
As such, an important ingredient for robustness of a grooming
or tagging algorithm is having a limited sensitivity to non-perturbative
contributions, such as hadronization or underlying event.
This robustness has already been demonstrated for the mMDT and SD
algorithms \cite{Dasgupta:2013ihk,Larkoski:2014wba}, and we present
here a similar analysis for RSD.

We use the process $pp \to Z+ j$ to generate samples of background QCD
jets, and use the
benchmark RSD parameters $\beta=1$ and $\zcut = 0.05$.
In \Fig{fig:m_NP_hadr}, we plot the ratio of the jet mass distributions
before and after the hadronization step in \pythia.
Without any grooming, there are around 10\% hadronization corrections throughout the whole distribution, with large corrections below a jet mass of $m_j\sim 50 \GeV$.
With RSD$_N$, though, the distribution is significantly more stable
down to jet masses of around 20 GeV, independent of the number of RSD
layers.
Remarkably, in the bulk of the distribution between 50--400 GeV,  RSD$_\infty$ exhibits around 5\% hadronization corrections.
At large mass, the RSD results also show a sizable improvement
as one increases the number of RSD layers. 

In \Fig{fig:m_NP_UE}, we show the impact of underlying event,
plotting the ratio of the jet mass distributions before and after the
inclusion of multiple parton interactions (MPI).
Here again, we see a similar behavior for all $N>0$ curves in the
small mass limit, with relatively small corrections due to non-perturbative
underlying event effects.
We observe furthermore that as $N\rightarrow\infty$, the stability of
the jet mass distributions improves substantially at high masses, such that the overall corrections are less than 10\% throughout the distribution.
It is therefore clear that RSD with $\beta>0$ substantially improves
the robustness of groomed jets to non-perturbative effects, notably by
providing more stable results than SD for large jet masses.
We also checked that the jet $p_t$ was stable to both hadronization and underlying event effects, with similar performance for all $N \ge 1$.

%======================================================================
\subsection{The $N\rightarrow\infty$ limit of zero-area jets}
\label{sec:zero-area}
%======================================================================

An interesting property of RSD groomed jets is that their catchment
area~\cite{Cacciari:2008gn} goes to zero in the $N \to \infty$
limit.
This is due to the fact that soft ghost particles with infinitesimal energy always fail the SD
condition~in \Eq{eq:SD-crit} for $\beta \geq 0$, such that the final
jets always have vanishing active and passive areas.%
\footnote{This zero-area feature is also shared by the semi-classical
  jet algorithm~\cite{Tseng:2013dva} and the ``priority'' jet
  algorithm~\cite{Duffty:2016dlg}.  A key difference is that RSD can
  be applied to standard anti-$k_t$ jets.}
For this reason, one might expect that RSD jets would be particularly
robust to pileup contamination, a feature we will explore further in
\Sec{sec:pileup-mitigation}.
That said, many of the commonly used pileup mitigation techniques,
such as the area--median method~\cite{Cacciari:2007fd,Cacciari:2008gn},
rely in some way on the jet area.
Applying these methods to zero-area jets, obtained when grooming with
$\RSDinf$, requires some care (see \Sec{sec:PU-area}).

\begin{figure}
  \centering
  \subfloat[]{%
    \includegraphics[width=0.45\textwidth]{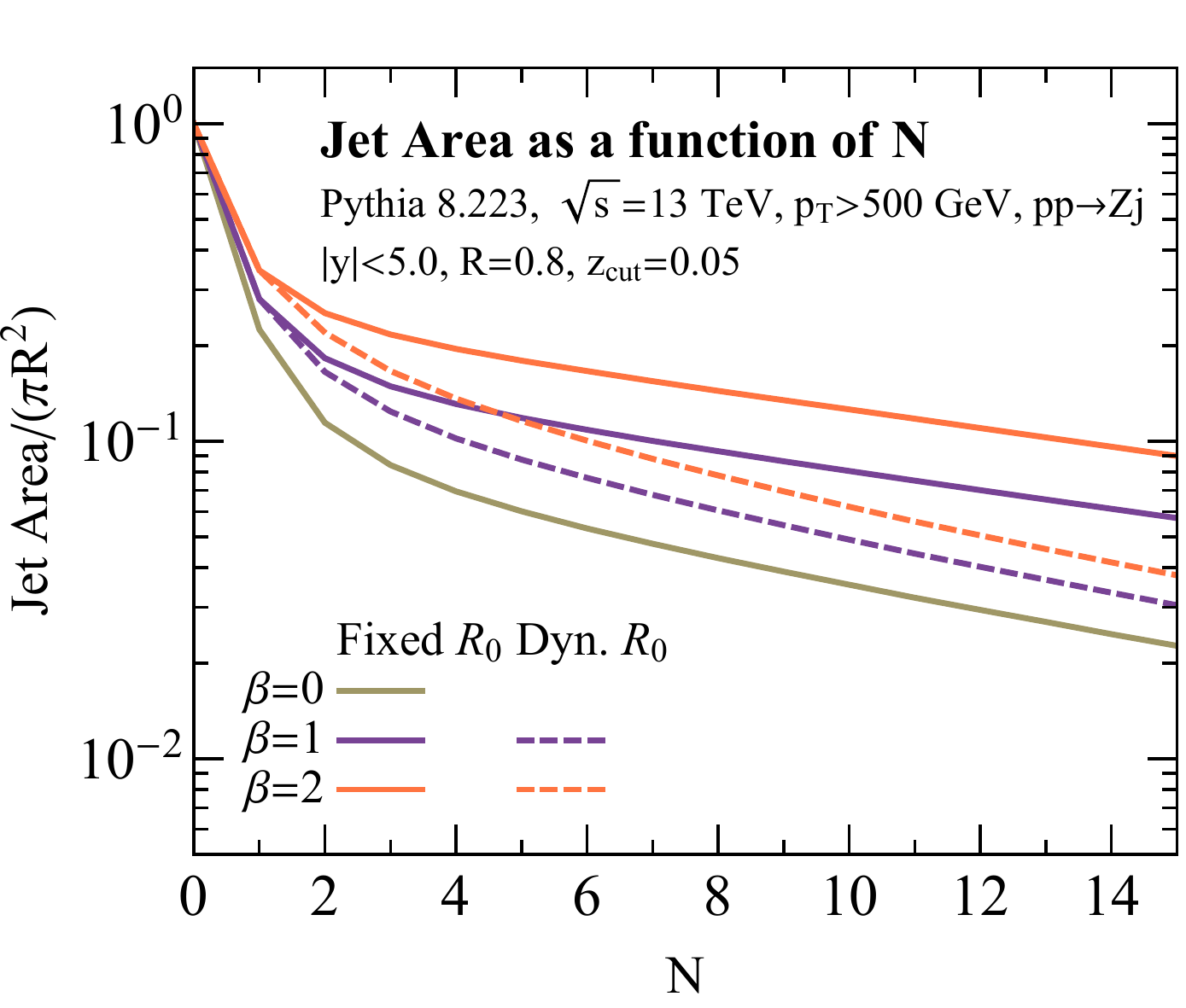}%
    \label{fig:jetarea_N}}
  \qquad
    \subfloat[]{%
    \includegraphics[width=0.45\textwidth]{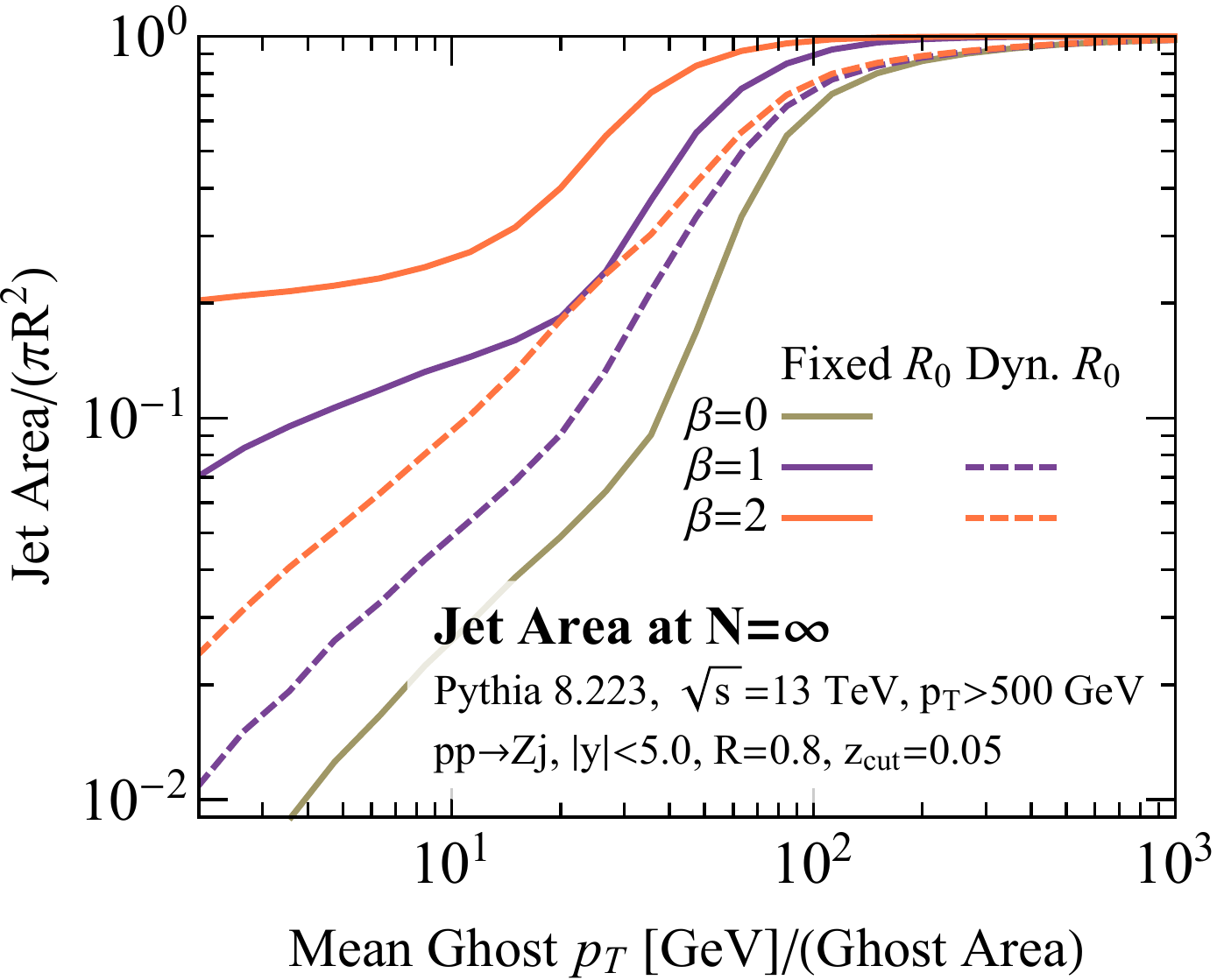}
    \label{fig:jetarea_rho}}
  \caption{Jet area studies on quark/gluon jets in $pp \to Zj$ events, for the fixed $R_0$ (solid line) and dynamic $R_0$
    (dashed line) algorithms.  (a) Active jet area as a function of the number of soft drop layers $N$.  (b) Active jet area as a function of
    ghost transverse momentum density $\rho$ for RSD$_\infty$.
    }
  \label{fig:jetarea}
\end{figure}

In \Fig{fig:jetarea_N}, we use the same $pp\rightarrow Z+j$ samples from \Sec{sec:NP-effects} and plot the active jet area as a
function of the number of RSD layers.
Here, we fix $\zcut = 0.05$ and scan the exponent $\beta = \{0,1,2\}$, and we explore both the fixed $R_0$ and dynamic $R_0$
variant from \Sec{sec:dynamicR}.
For all choices of $\beta$, the active jet area decreases exponentially with
$N$, as anticipated. 
For smaller $\beta$, the algorithm is more aggressive at
removing soft radiation, such that the jet area decreases the most rapidly for
$\beta=0$.
In \Fig{fig:jetarea_N}, one can see that the dynamic $R_0$
algorithm yields jets with smaller active area for any given $\beta$
value.
In this way, dynamic $R_0$ behaves more closely to the $\beta=0$
limit of the fixed $R_0$ algorithm, leading to decreased pileup
sensitivity.

Although RSD$_\infty$ leads to zero-area jets in a formal sense, soft
particles from, say, pileup have finite energy in an experimental
setting.
In \Fig{fig:jetarea_rho}, we plot the effective jet area for $\RSDinf$
with finite-energy ghost particles, again considering
$\beta = \{0,1,2\}$.
Here, we report the ghost transverse momentum flow density per unit
area, which is roughly 0.5 GeV per $1.0\times1.0$ bin in
rapidity/azimuth for one minimum bias collision.
For transverse momentum flow densities starting around 50 GeV---below the pileup $p_t$ densities anticipated at the
high-luminosity LHC~\cite{Aad:2015ina}---we observe that while the
jet area after RSD$_\infty$ grooming is reduced, it remains
substantially above zero even in the $N=\infty$ limit.
This behavior explains in part why RSD is not sufficient in itself to remove pileup, and instead performs best when combined
with another pileup mitigation technique, as discussed in~\Sec{sec:pileup-mitigation}.

%======================================================================
\section{Improved mass resolution}
\label{sec:mass-resol}
%======================================================================

The simplest way to identify a boosted hadronically-decaying resonance is through the invariant mass of its decay
products; in a contamination-free setting, this would correspond to the plain jet mass.
In practice, though, the jet mass is particularly sensitive to
unassociated soft wide-angle emissions which smears out the
distribution.
To restore the mass resolution, it is therefore necessary to mitigate soft contamination through appropriate grooming.

While mMDT and SD are not always the most effective methods
for enhancing the signal efficiency for boosted objects, they have the advantage of being
particularly robust~\cite{Dasgupta:2016ktv}.
As seen in \Sec{sec:NP-effects}, hadronization and underlying event
effects are suppressed, and they have an analytically-tractable
behavior due to the absence of double logarithms and leading
non-global
contributions~\cite{Dasgupta:2001sh,Dasgupta:2002bw,Dasgupta:2013ihk}.
To maximize the tagging performance, though, \Ref{Butterworth:2008iy} found that MDT grooming should be supplemented by an extra filtering step to improve the jet mass resolution.
The hope is that an algorithm like RSD$_\infty$ could, by extending
the grooming procedure down to smaller angular scales, achieve
excellent mass resolution without requiring any further
post-processing.

In this section, we study the mass resolution with RSD in three cases
of interest for the LHC: 2-prong boosted $W$ bosons, 3-prong boosted
top quarks, and 4-prong boosted Higgs jets ($H\to VV\to 4f$).
In each case, we use the same \textsc{Pythia} 8.223 generator settings
from \Sec{sec:basic}, and consider all jets in the event
that pass the selection cuts $p_t>500 \GeV$ and $|y| < 5$.
As we will see, RSD provides a way to improve the achievable
resolution, with gains in the 10-20\% range, while retaining the
tractability and robustness of SD.

Our overall recommendation from these studies will be to use
RSD$_\infty$ with the default settings of $\beta = 1$ and
$\zcut =0.05$, which gives good performance across the three test
cases.  (although other values of $\beta$ and $\zcut$ are worth
investigating as well).
This conclusion will persist even after the inclusion of pileup in \Sec{sec:mass_PU}, making RSD$_{\infty}$ useful in extreme environments such as the one faced by the high-luminosity LHC.

%======================================================================
\subsection{Definition of the mass peak and resolution}
\label{sec:bestfit}
%======================================================================

To define the mass resolution, we identify the \emph{smallest} mass
interval that contains a fixed fraction $f$ of the total event
samples (see e.g.~\cite{Altheimer:2013yza}).
For concreteness, we set $f = 0.4$ for these studies.
The central value is then defined as the median of the mass interval, and the
width is defined as the width of the mass interval.

An alternative approach would be to fit the mass distribution with two curves, a narrow Cauchy distribution to capture the signal and a wider background distribution to account for cases of poor reconstruction.
The advantage of the interval method is that it allows us to avoid
biases associated with the choice of a fitting functional form.
That said, we did test the fitting approach and got similar qualitative features to the ones shown here.
We also tested that the choice of fraction $f$ did not affect the qualitative conclusions.

One
caveat of the interval method is that it combines two
logically distinct effects: mass resolution and signal efficiency.
For example, if a technique yields perfect mass resolution, but only on a small subsample of events, then the mass interval will be larger than the resolution in order to include the fixed fraction $f$.
In practice, though, many boosted-object taggers select jets based on a fixed mass window, so our definition of mass resolution is appropriate for that setting.

%======================================================================
\subsection{Two-prong $W$ decays}
\label{sec:W_mass}
%======================================================================

\begin{figure}[t]
  \centering
  \subfloat[]{%
    \includegraphics[trim={0 0 0 0.5cm},clip,width=0.48\textwidth]{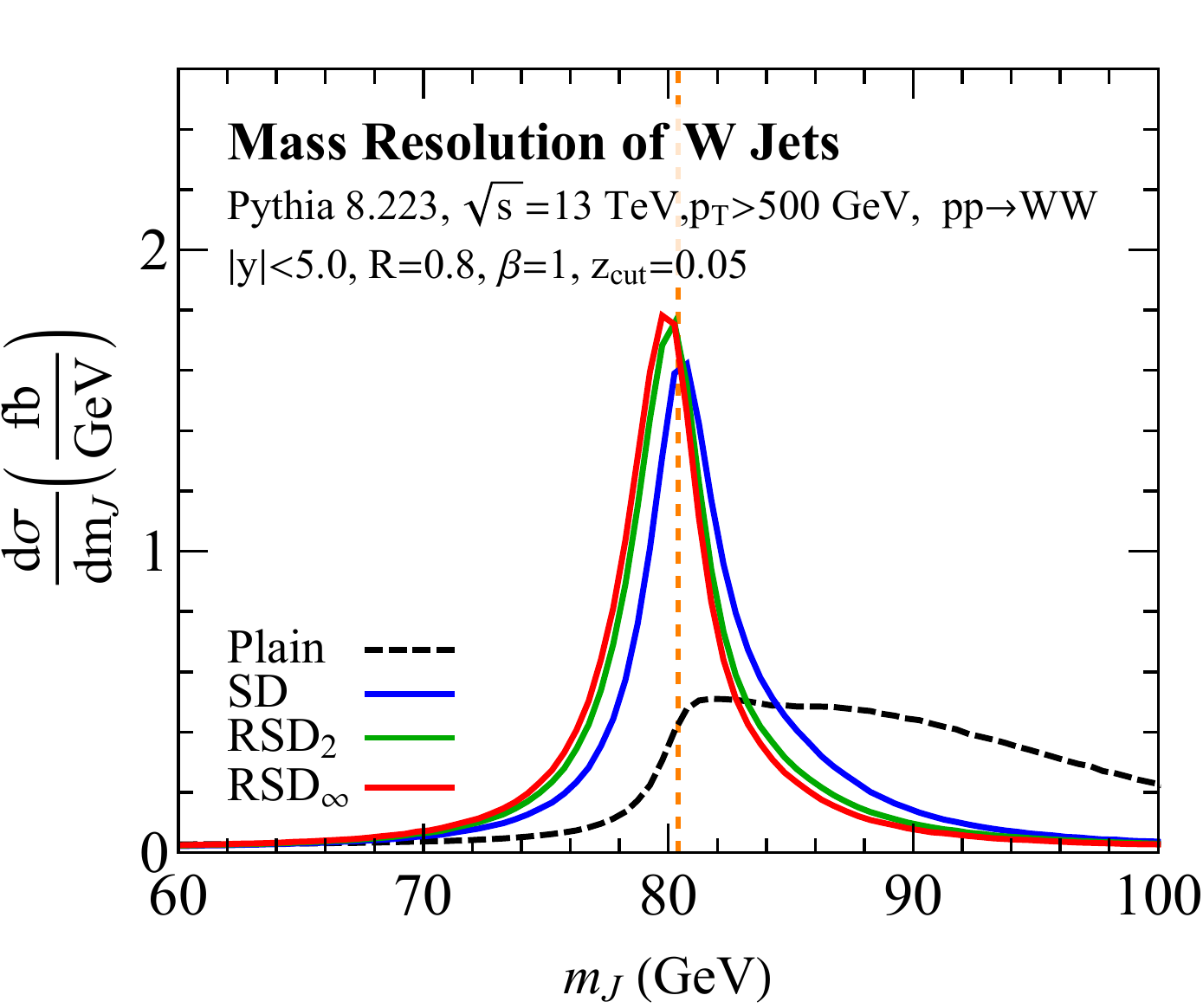}%
    \label{fig:W_mass_distribution_dist}}%
  \qquad
  \subfloat[]{\label{fig:W_mass_distribution_fit}%
  \begin{tabular}[b]{c}
    \includegraphics[trim={0 1.51cm 0 0},clip,width=0.4\textwidth]{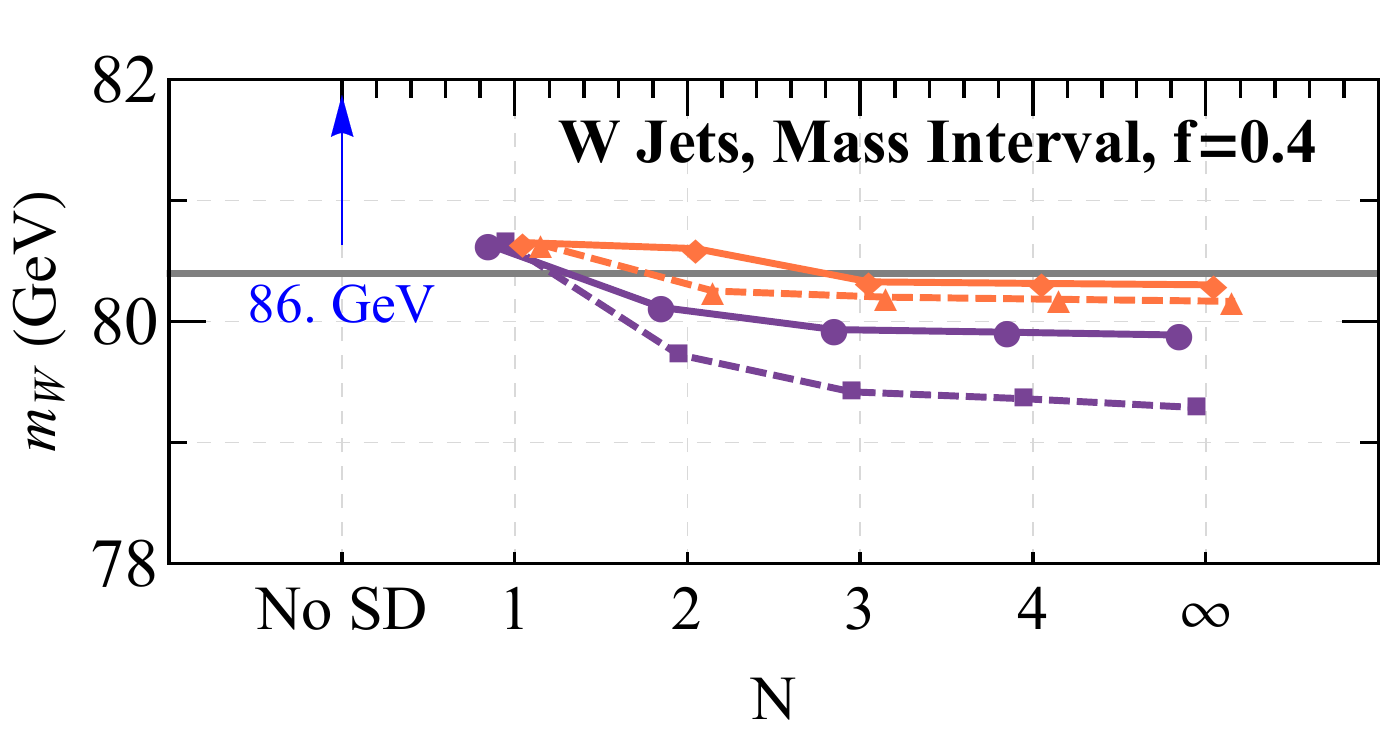} \\[2mm] % 
    \includegraphics[trim={-0.35cm 0 0 0.5cm},clip,width=0.4\textwidth]{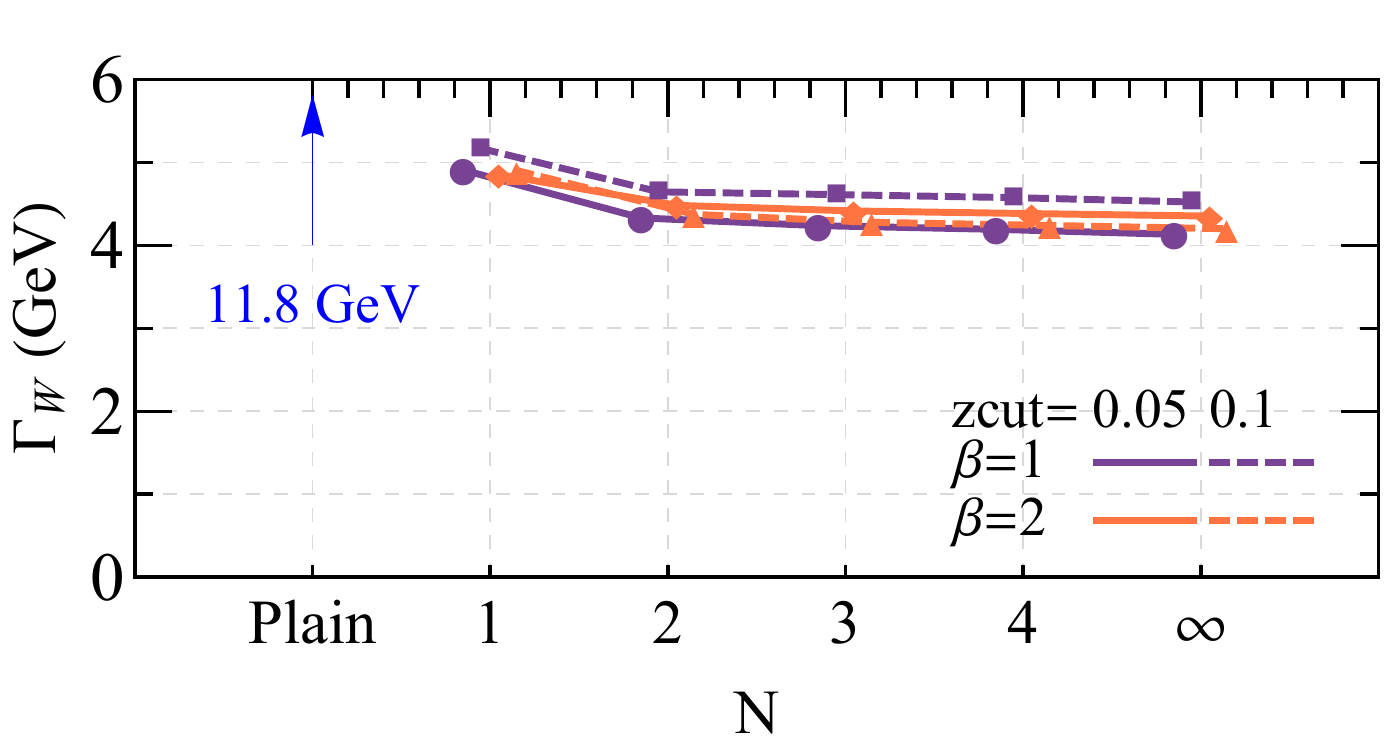}%
    \end{tabular}}
  \caption{(a) Mass distributions for the $W$ jet sample using different levels of RSD and the benchmark values of $\beta = 1$ and $\zcut =0.05$.  (b) Central mass values and widths as a function of $N$, testing four different $\beta$ and $\zcut$ combinations, using a mass interval containing a fraction $f=0.4$ of the events.
  }
  \label{fig:W_mass_distribution}
\end{figure}

In \Fig{fig:W_mass_distribution_dist}, we show the jet mass distribution from a $W$ jet sample obtained from $pp \to WW$, with the $W$ decaying hadronically.
At least some level of grooming is required to obtain a peak within $10\%$ of the $W$ mass.
With $\beta = 1$ and $\zcut =0.05$, the jet mass distribution is already close to the expected $W$ mass value after applying SD alone.
Adding additional SD layers with $N \geq 2$, the peak shifts somewhat
below the expected $W$ mass value, but, more importantly, the width of
the mass distribution decreases.
Another interesting observation is that
the mass distribution is more symmetric 
with additional grooming layers, such that with $N=\infty$, the jet
mass can be accurately fit by a Cauchy distribution.

To get a sense of the dependence on the choice of RSD parameters, in \Fig{fig:W_mass_distribution_fit} we plot the central value and width of the mass distribution as a function of $N$.
As discussed in \Sec{sec:bestfit}, we use a mass interval containing a fraction $f=0.4$ of events.
The benchmark parameters of $\beta=1$ and  $z_{\text{cut}}=0.05$ undershoot the $W$ mass central value by around 1 GeV, while switching to $\beta=2$ gives a better reconstruction of the central value.
On the other hand, the benchmark parameters yield the best $W$ mass resolution, which implies somewhat better tagging performance.
More generally, all of the RSD settings yield a sizable improvement
over the ungroomed case and a smaller but clearly visible improvement,
of order 10--20\%, over the SD case in terms of resolution.

%======================================================================
\subsection{Three-prong top decays}
\label{sec:top_mass}
%======================================================================

\begin{figure}[t]
  \centering
    \subfloat[]{%
    \includegraphics[trim={0 0 0 0.5cm},clip,width=0.47\textwidth]{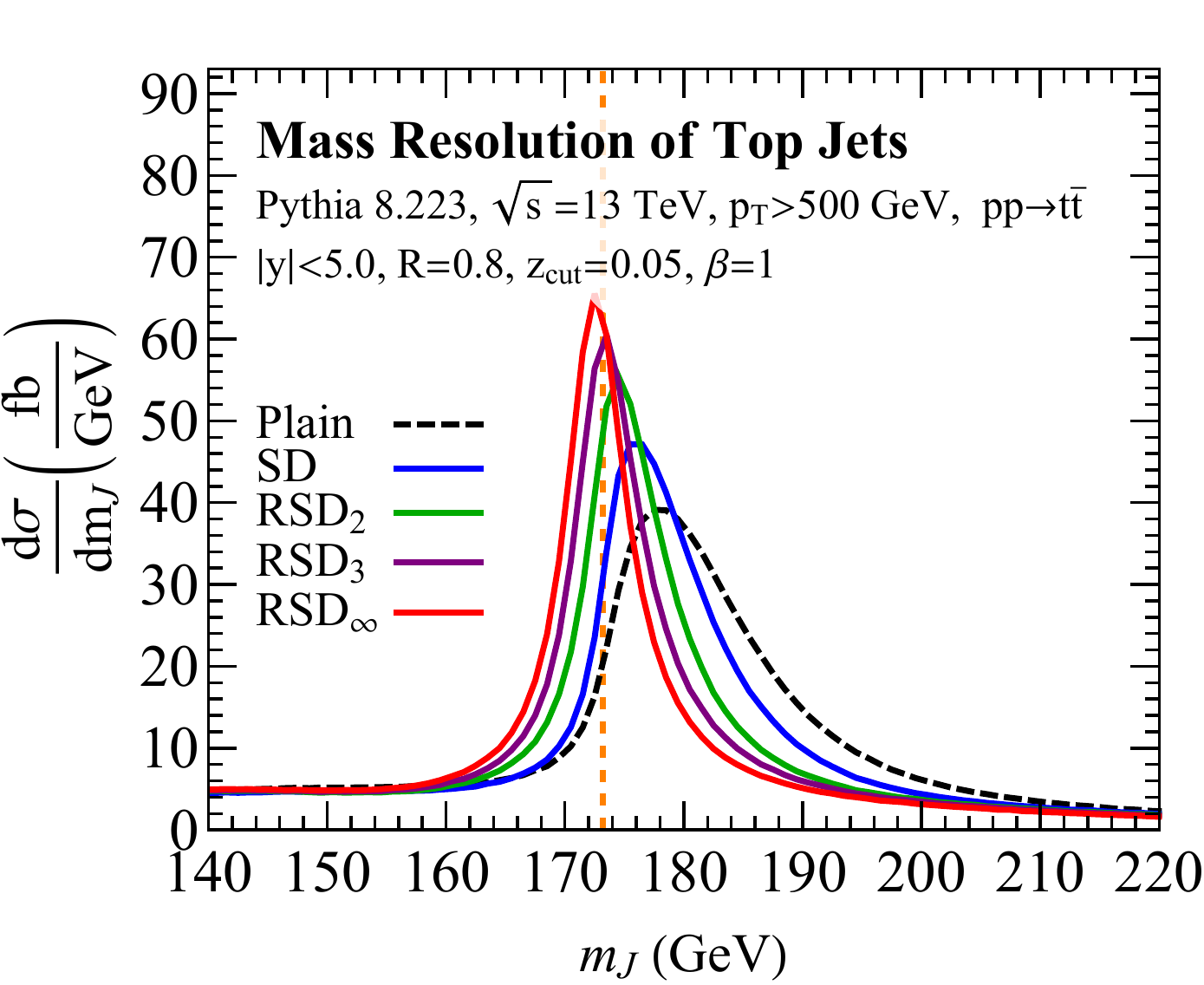}%
    \label{fig:top_mass_distribution_dist}}%
  \qquad
  \subfloat[]{\label{fig:top_mass_distribution_fit}%
  \begin{tabular}[b]{c}
    \includegraphics[trim={0 1.51cm 0 0},clip,width=0.4\textwidth]{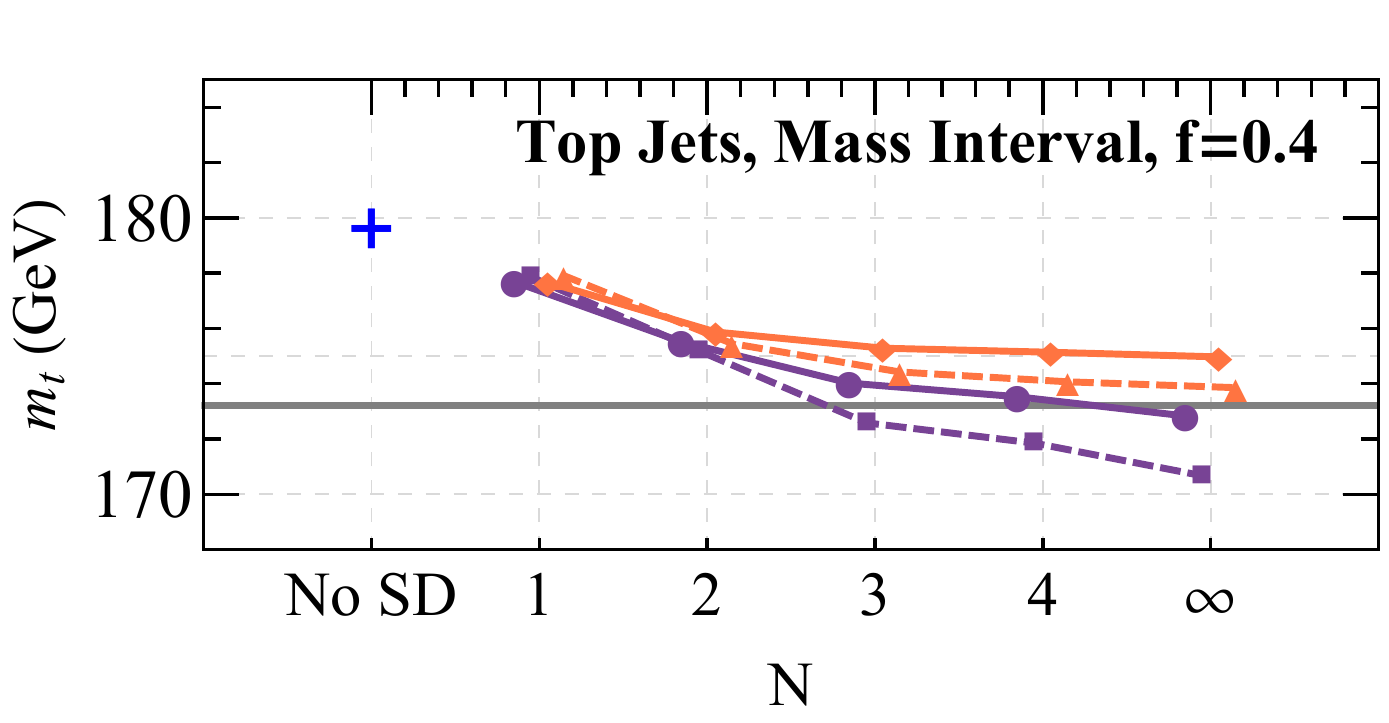} \\[2mm] % 
    \includegraphics[trim={-0.4cm 0 0 0.5cm},clip,width=0.4\textwidth]{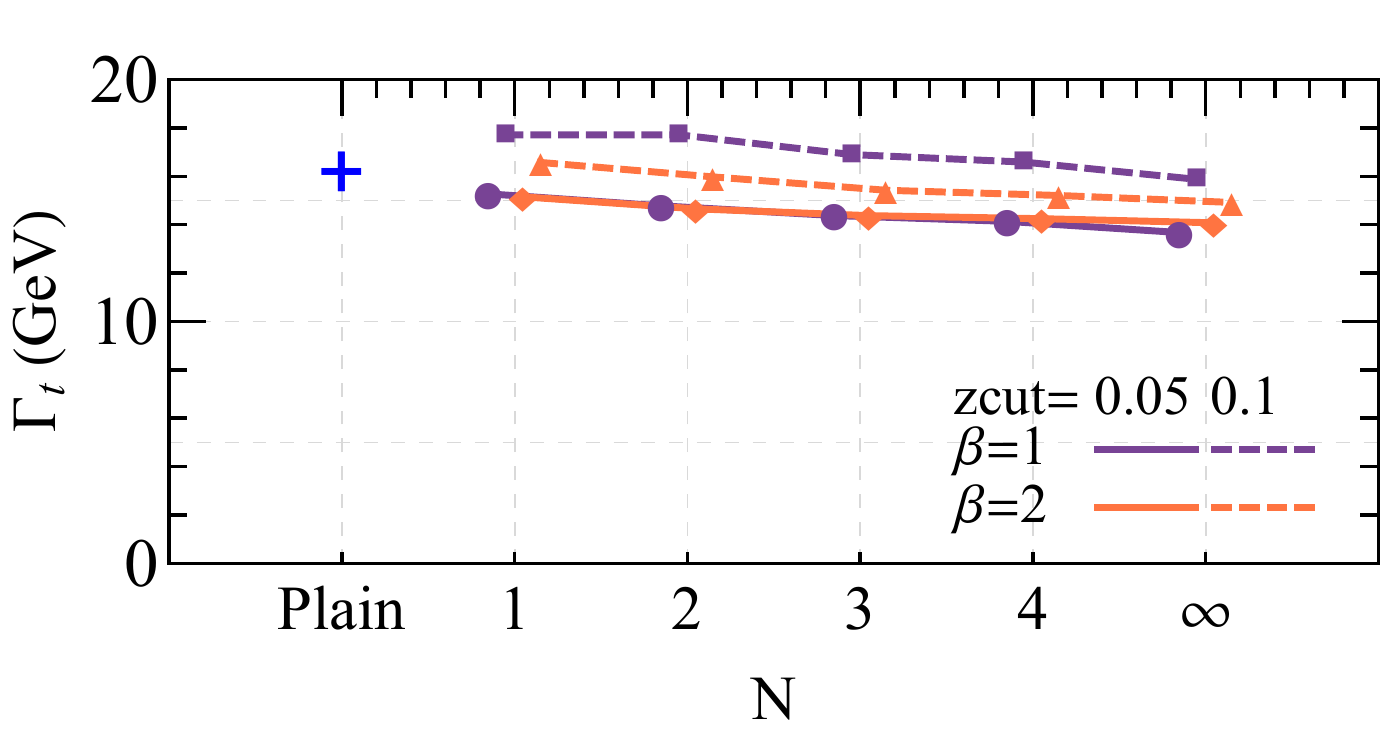}%
    \end{tabular}}
  \caption{ Same as \Fig{fig:W_mass_distribution} but for the top jet sample.}
  \label{fig:top_mass_distribution}
\end{figure}

Unlike SD, which terminates after probing only the leading two subjets, RSD with $N > 1$ is well suited to handle the broad radiation patterns and three-prong topologies of top quarks.
We consider a sample of $pp \rightarrow t\overline{t}$ events, where the top mass in \pythia{} is set to 173.2 GeV.
In \Fig{fig:top_mass_distribution_dist}, we show the mass distribution of boosted tops for different layers of RSD.
As expected, we find that without grooming, the peak of the
distribution occurs at values well above the top mass.
With successive layers of RSD grooming, the mass peak shifts to
lower values, and converges very close
to the top mass for $N=\infty$.
This is due to the fact that we discard more of the extra
unassociated radiation, and therefore the mass tends to decrease.

In \Fig{fig:top_mass_distribution_fit}, we consider the central mass value and width containing a fraction $f=0.4$ of the mass distribution.
We find that $\beta=1$ and $z_{\rm{cut}} =0.05$ gives nearly optimal
performance (by the mass interval measure), with $\beta=2$ and
$z_{\rm{cut}} =0.1$ giving comparable performance in terms of accuracy
and resolution.
The gain in resolution compared to SD is found to be slightly larger
than 10\%.

Compared to the $W$ case in \Sec{sec:W_mass}, where the resolution tends to saturate by $N = 2$, the top case benefits from taking $N \geq 3$.
In fact, there is a marginal benefit from taking $N \to \infty$, which is why we recommend RSD$_\infty$ for mass resolution studies involving the top quark.
It would be interesting to see whether RSD$_{\infty}$ would also be appropriate for defining a short-distance top quark mass \cite{Hoang:2017kmk}.

%======================================================================
\subsection{Four-prong Higgs decays in associated production}\label{sec:higgs_mass}
%======================================================================

\begin{figure}[t]
\centering
\subfloat[]{
\label{fig:H_mass_distribution_dist}
\includegraphics[trim={0 0 0 0.5cm},clip,width=0.47\textwidth]{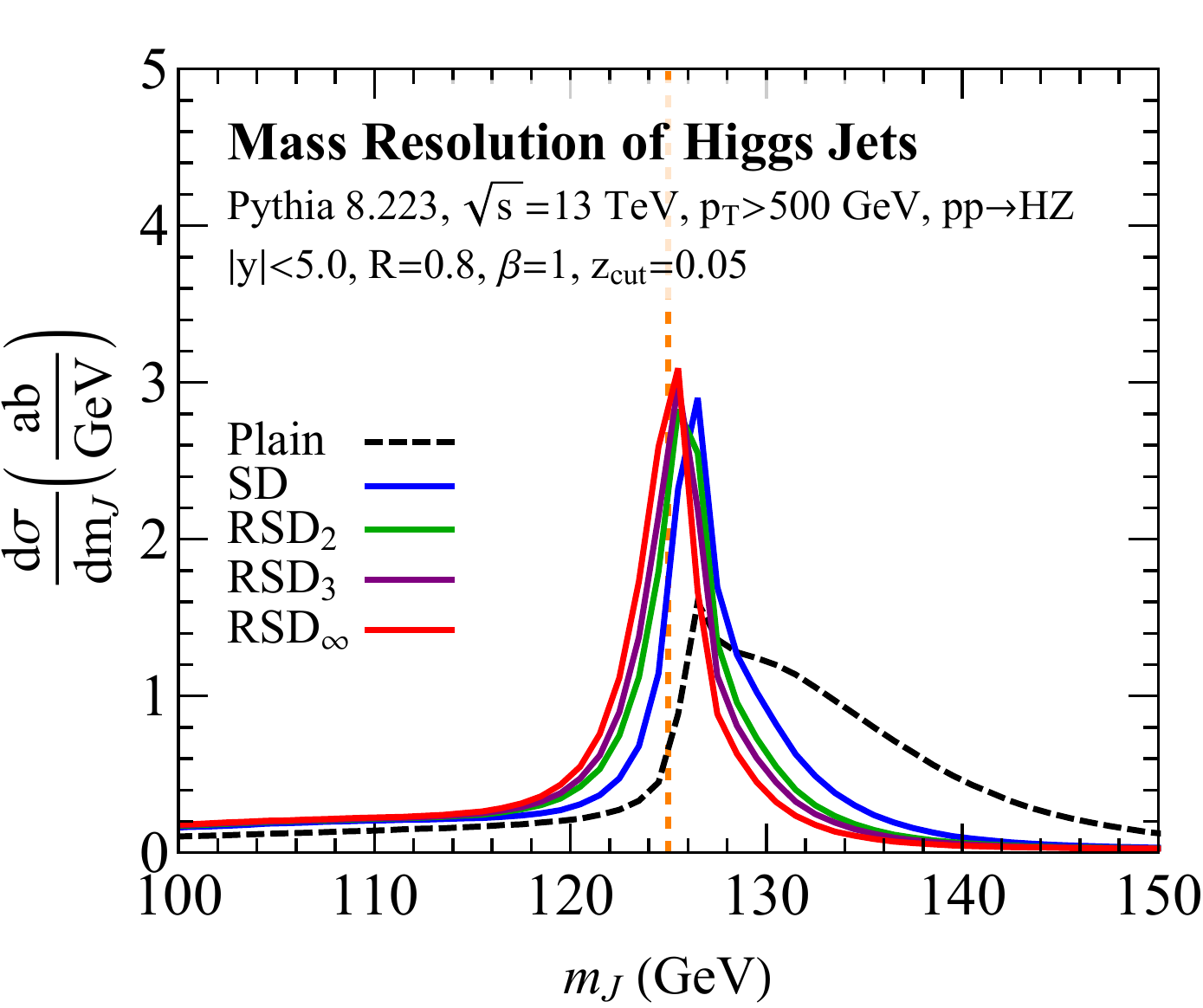}
}
\qquad
\subfloat[]{\label{fig:H_mass_distribution_fit}
\begin{tabular}[b]{c}
\includegraphics[trim={0 1.51cm 0 0},clip,width=0.4\textwidth]{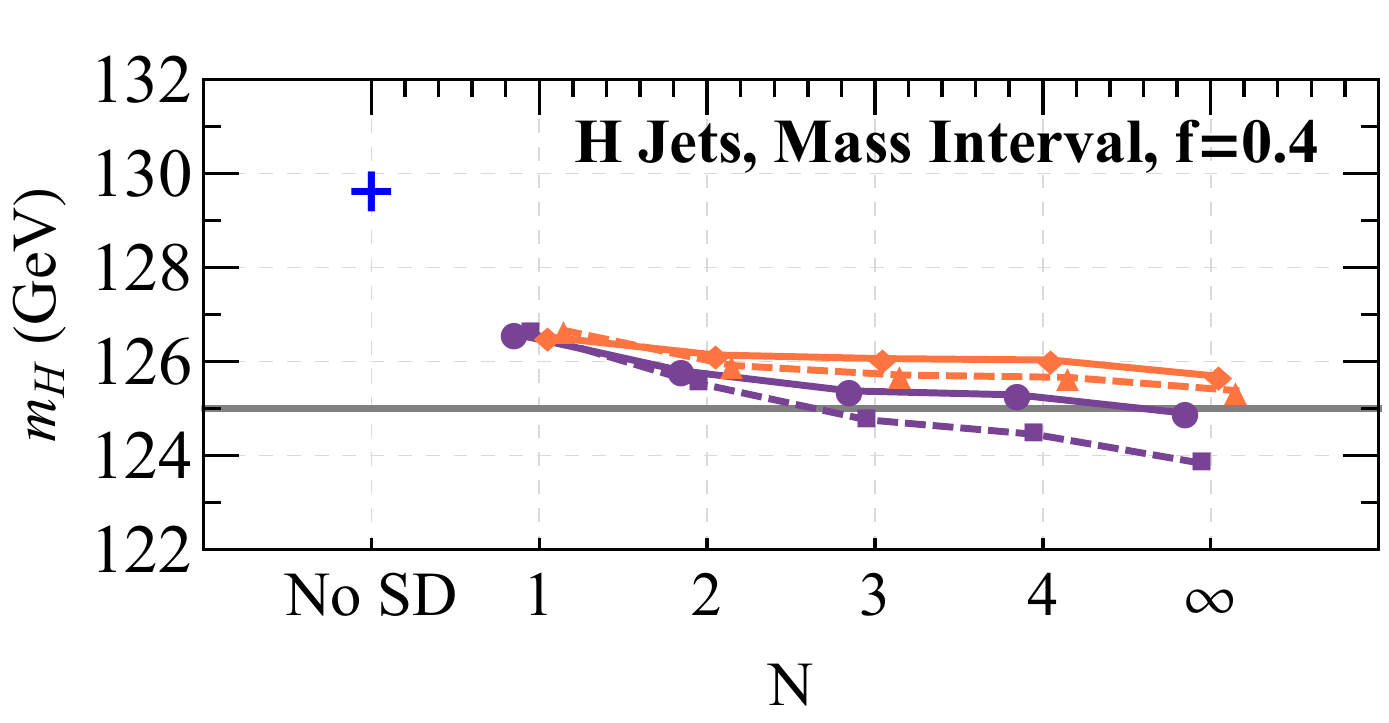} \\[2mm] % 
\includegraphics[trim={-0.4cm 0 0 0.5cm},clip,width=0.4\textwidth]{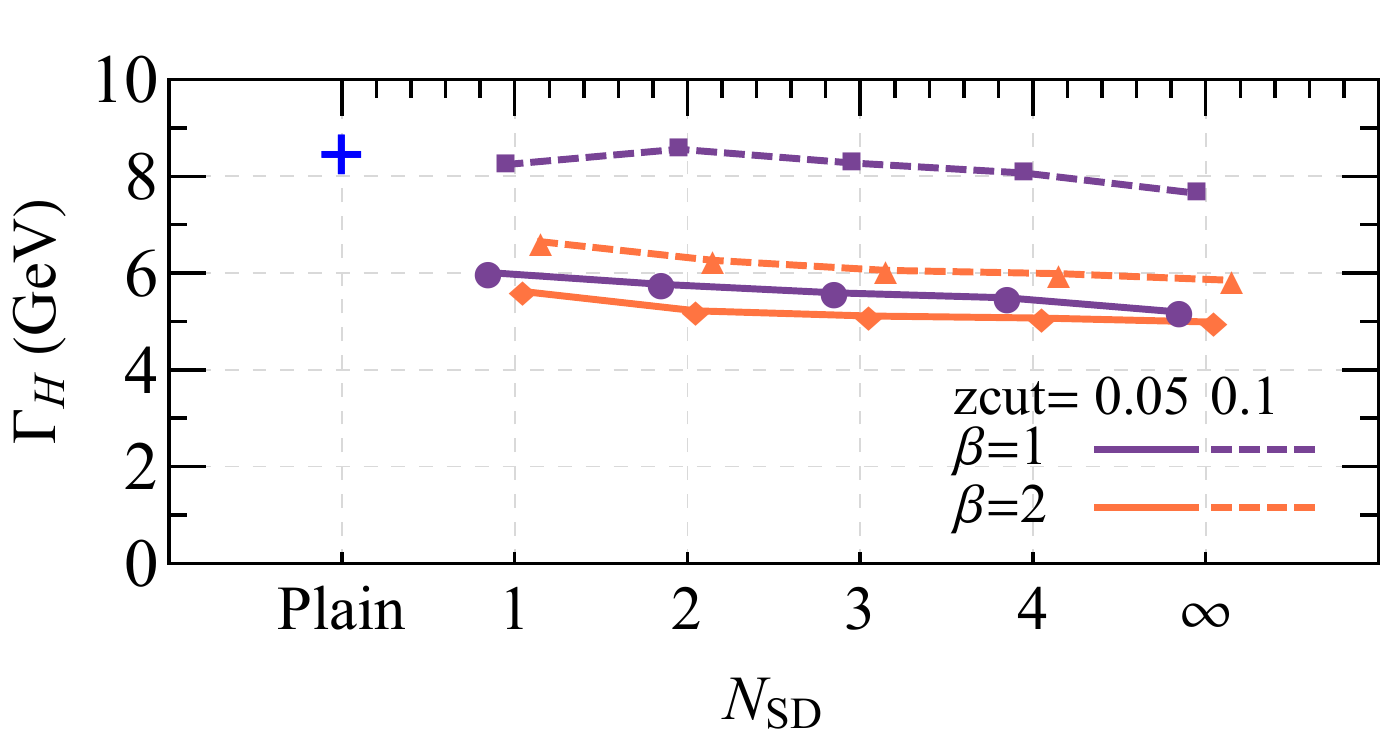}%
\end{tabular}
}
\caption{Same as \Fig{fig:W_mass_distribution} but for the Higgs jet sample.}
\label{fig:H_mass_bestfit_distribution}
\end{figure}

Four-prong jets can provide important information in certain Higgs
decays, where the $H\rightarrow VV\rightarrow 4f$ channel plays an
important role in determining properties of the
Higgs~\cite{Choi:2002jk} as well as its couplings to
bosons~\cite{Meyer:2004ha}.
They can also provide a useful probe for new physics, such as in
hadronic decays of heavy resonances decaying to $HW$ or
$HZ$~\cite{Khachatryan:2015bma}, or in hadronic decays of a boosted
Higgs pair~\cite{Bishara:2016kjn}.

In \Fig{fig:H_mass_distribution_dist}, we show the reconstructed Higgs mass in boosted $H\rightarrow W^+W^-\rightarrow q\bar{q}q\bar{q}$ decays.
This comes from a sample of $pp \rightarrow HZ$ events, with the $Z$
boson decaying to neutrinos.
Although grooming with ordinary SD performs relatively well,
the Higgs mass reconstruction is better with $N\geq 3$,
as one would expect for a four-particle decay.
A comparison of the central mass values and width for different
grooming parameters is shown in \Fig{fig:H_mass_distribution_fit},
showing gains in resolution around 10--15\%.
Once again, the benchmark choice of $\beta=1$ and $z_{\rm{cut}} =0.05$
gives an excellent reconstruction, especially in the $N \to \infty$
limit, although $\beta=2$ and $z_{\rm{cut}} =0.05$ gives a slightly
better resolution.

%======================================================================
\subsection{Boosted top tagging}
\label{sec:boosted-toptag}
%======================================================================

We conclude this section with a concrete example of the impact of improved mass resolution from RSD.
Using the same samples as \Sec{sec:top_mass}, we perform a study of boosted top tagging performance.
We consider two standard observables used for discriminating top jets from QCD background:
the $N$-subjettiness ratio
$\tau_{32}=\tau_3/\tau_2$~\cite{Thaler:2010tr,Thaler:2011gf}, and the generalized
energy correlation function ratio $N_3^{(2)} = {}_2e^{(2)}_{4}/({}_1e^{(2)}_3)^2$~\cite{Larkoski:2013eya,Moult:2016cvt}.
By adjusting the degree of grooming---on both the jet mass and on the jet discriminant---we can assess the potential performance gains from RSD.

\begin{figure}
  \centering
  \includegraphics[width=0.5\textwidth]{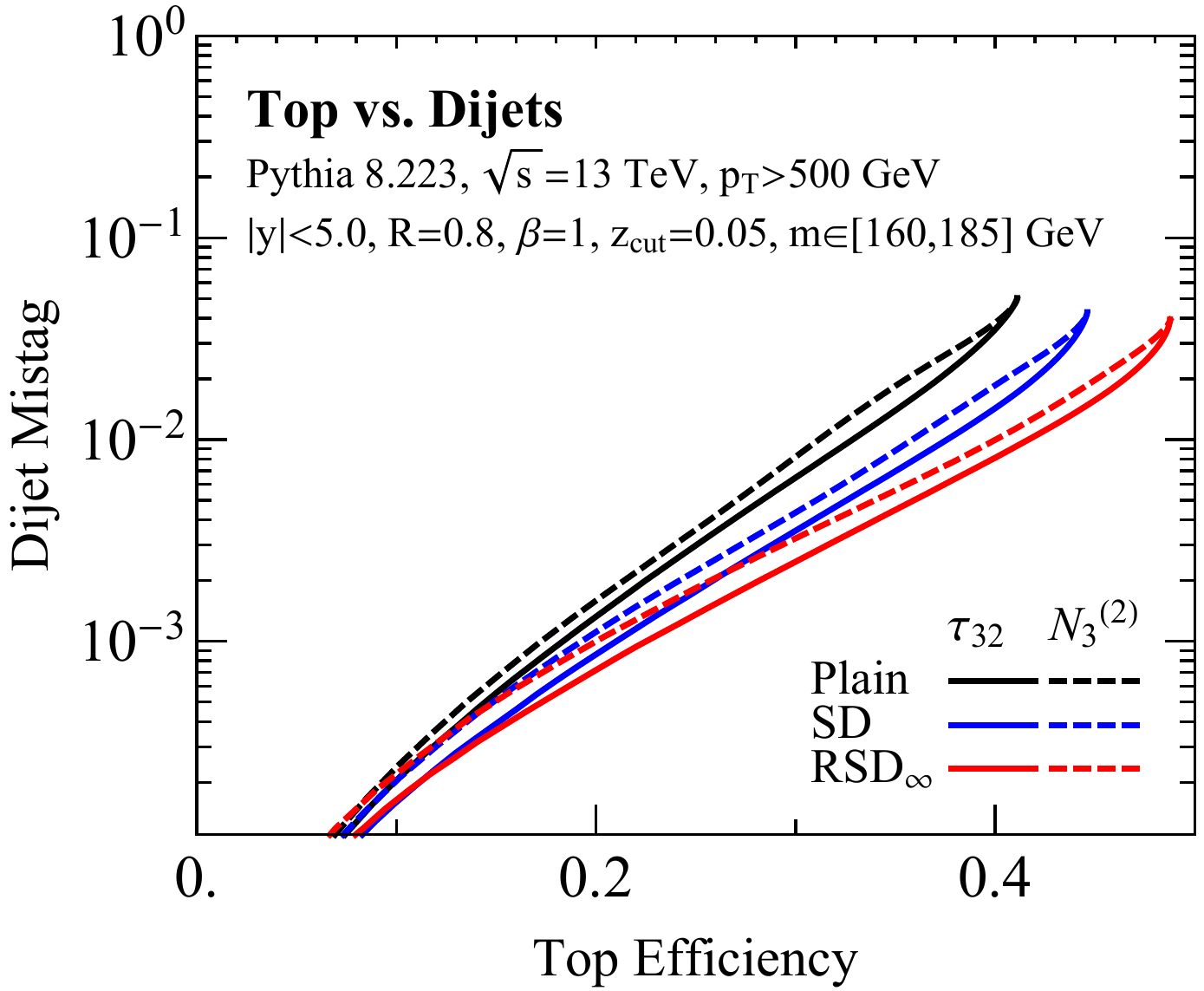}
  \caption{Top jet signal efficiency versus dijet background mistag rate, with the benchmark RSD parameters of $\beta=1$ and $\zcut=0.05$.}
  \label{fig:discrim-ROC}
\end{figure}

In~\Fig{fig:discrim-ROC}, we plot the top signal efficiency versus dijet background mistag rate (ROC curves).
The rightmost endpoints of the ROC curve indicate the impact of an overall cut of $m_{\text{groomed}}\in[160,185]$.
As $N$ increases, the dijet mistag rate improves somewhat, and the
top signal efficiency increases due to the improved mass
resolution.
The rest of the ROC curve is associated with sweeping a cut on $\tau_{32}$ and $N_3$.
Note that as the substructure cut becomes more stringent, there is less of a gain from increasing the number of grooming layers.
The reason is that grooming removes part of the radiation phase-space where there is still some discriminating power, or equivalently, a cut on $\tau_{32}$ or $N_3$ already removes some of the phase-space regions targeted by grooming.
Because this phase-space region is also the most sensitive to non-perturbative effects, there is a tradeoff between tagging performance and non-perturbative robustness~\cite{Dasgupta:2016ktv}.
Despite this tradeoff, RSD$_\infty$ maintains the best tagging performance for top efficiencies greater than 10\%.

As discussed in \Ref{Moult:2016cvt}, $\tau_{32}$ with grooming and $N_3$ with grooming have the same soft-collinear power counting, so their performance is expected to be similar (see further discussion in \Ref{Larkoski:2014gra}).
This conclusion persists even with multiple grooming layers.
The relative difference between $\tau_{32}$ and $N_3$ then depends on the precise details of the event selection.
$N_3$ is not defined with respect to any axes, so it is expected to perform better than $\tau_{32}$ in kinematic regimes where the choice of axes can be ambiguous \cite{Moult:2016cvt}.
Here, though, we are taking a rather tight mass window of $m_{\rm jet}
\in [160,185]$ GeV, so the axes effects are subleading and $\tau_{32}$
turns out to give better performance.
Though not shown, we also tested the performance of the $M_3$ observable \cite{Moult:2016cvt}.
While $M_3$ always performs worse than $\tau_{32}$ or $N_3$, the relative improvement in going from SD to RSD$_\infty$ is greater due to
the removal of radiation in all three prongs.

We found similar tagging results for other boosted objects, such as hadronic decays of boosted $W$ and Higgs bosons.
In general, RSD$_\infty$ gives similar or improved performance compared to tagging with SD, though most of that comes just from the improved mass resolution.
For the case of $W$ tagging,
  it would be interesting to study a version of the dichroic
  $N$-subjettiness ratio~\cite{Salam:2016yht}, where RSD$_\infty$ is
  used to compute the jet mass and $\tau_1$ (or ${}_1e_2$), and a lighter
  grooming, like (R)SD with larger $\beta$ and smaller $\zcut$,
  is used to calculate $\tau_2$ (or ${}_2e_3$).
Results for boosted top tagging with large pileup multiplicity are shown in \App{sec:toptag-pileup}.

%======================================================================
\section{Robust pileup mitigation}
\label{sec:pileup-mitigation}
%======================================================================

As the LHC progresses towards ever higher luminosities, mitigating the
effect of secondary proton-proton collisions becomes an
increasingly important challenge.
Large pileup levels can substantially impact typical jet observables,
increasing for example the jet's momentum, while shifting and
distorting jet shapes.
A number of techniques have been developed to remove soft
radiation associated with secondary vertices.
These include area--median~\cite{Cacciari:2007fd,Cacciari:2008gn} or
shape~\cite{Soyez:2012hv} subtraction; grooming with
trimming~\cite{Krohn:2009th}, pruning~\cite{Ellis:2009me}, or mMDT/SD
\cite{Dasgupta:2013ihk,Larkoski:2014wba}; charged hadron
subtraction~\cite{CMS:2009nxa}; particle-level removal methods
such as SoftKiller~\cite{Cacciari:2014gra} and
PUPPI~\cite{Bertolini:2014bba}; and machine learning
methods~\cite{Komiske:2017ubm}.
In practice, in the context of jet substructure, it is generally most
useful to combine a removal or subtraction method with some type of
jet grooming to achieve optimal results (see
e.g.~\cite{Altheimer:2013yza}).

Here, we investigate the properties of jets after grooming with
RSD under varying levels of pileup, with additional pileup studies appearing in \App{app:additionalpileup}.
We will see that, when used in combination with SoftKiller
\cite{Cacciari:2014gra}, the RSD algorithm yields robust pileup mitigation
results even under high pileup conditions.%
\footnote{Applying RSD without any pileup mitigation shows poor performance, so we have not included those results in our discussion.}
We will also test RSD with the area--median subtraction approach~\cite{Cacciari:2007fd}.

%======================================================================
\subsection{Mass resolution with SoftKiller}
\label{sec:mass_PU}
%======================================================================

SoftKiller is an event-wide particle-level removal method, which uses
a dynamic cut on transverse momentum $p_t^{\text{cut}}$ to remove soft
particles~\cite{Cacciari:2014gra}.
The threshold is determined dynamically on an event-by-event basis as
follows:
\begin{enumerate}
\item The event is split into patches of size $a_{\rm SK} \times a_{\rm SK}$ in rapidity-azimuth space.
\item For each patch, one determines $p_{t,i}^{\text{max}}$, the largest particle transverse momentum in patch $i$.
\item The transverse momentum threshold $p_t^{\text{cut}}$ is determined by
\begin{equation}
  p_t^\text{cut} = \underset{i}{\text{median}}\, \left(p_{t,i}^{\text{max}}\right)\,,
\end{equation}
where the median is taken across all patches.
\item All particles with transverse momenta below $p_t^{\text{cut}}$ are removed from the event.
\end{enumerate} 
An equivalent description of SoftKiller is finding the minimal $p_t^{\text{cut}}$ such that
exactly half of the patches have zero momenta.
This $p_t^{\text{cut}}$ ensures that the median across patches of transverse-momentum flow
per unit area $\rho$ is zero.\footnote{The quantity $\rho$ is computed via $\rho = \underset{i}{\text{median}}\,\left( \frac{p_{t,i}}{A_i}\right)$,
where $A_i$ is the patch area and $p_{t,i}$ is the transverse momentum of the patch.}

\begin{figure}
  \centering
  \includegraphics[width=0.5\textwidth]{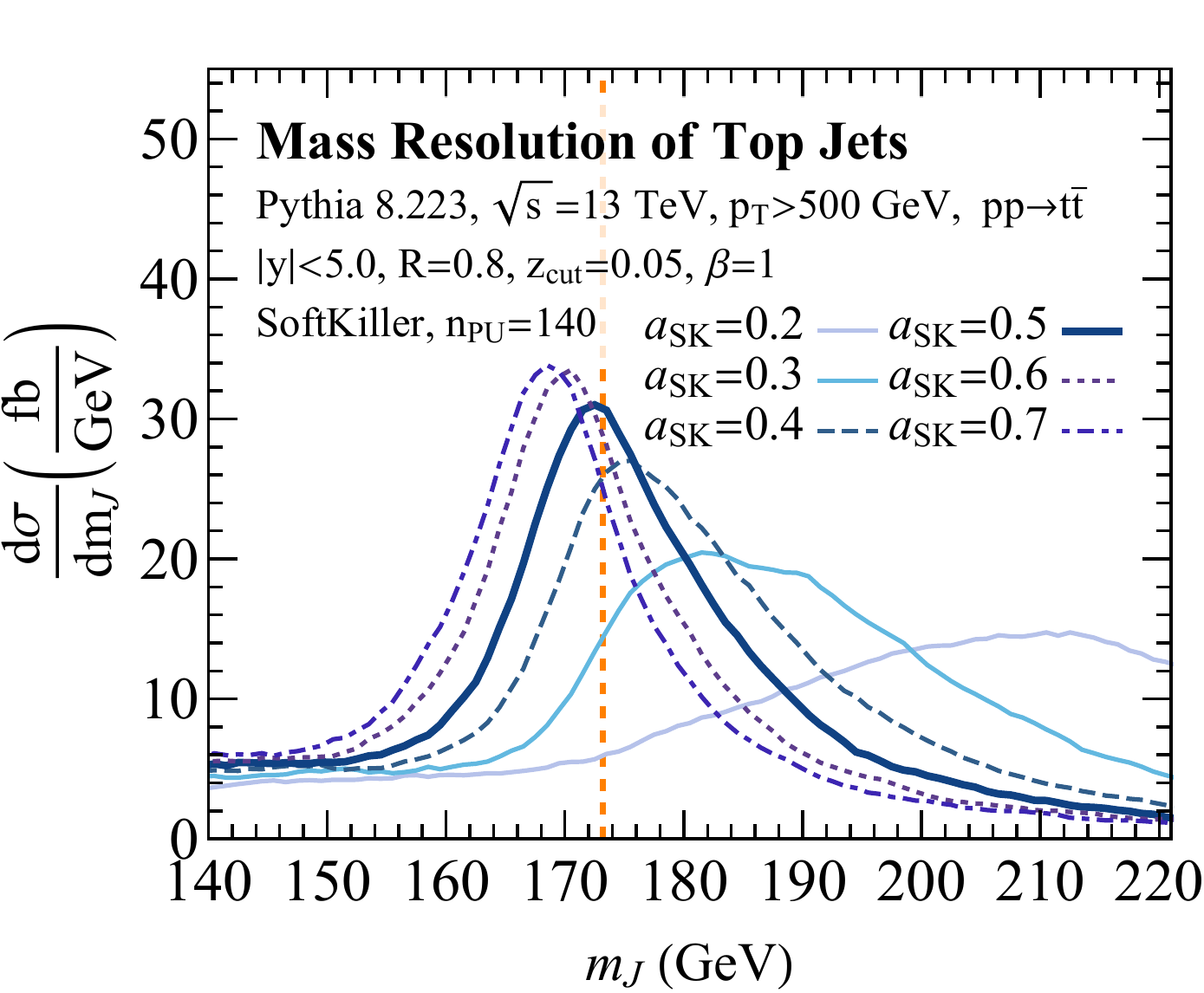}
  \caption{The RSD groomed mass distribution for multiple values of
    the SK grid parameter $a_{\rm SK}$.
    }
  \label{fig:SK-scan}
\end{figure}

For each analysis, one has to set the appropriate patch size by choosing the SoftKiller parameter $a_{\rm SK}$.
The goal is to set the parameter $a_{\rm SK}$ to minimize both the shift in jet $p_t$ and
mass.
After RSD grooming of $R=0.8$ jets, this usually achieved
with $a_{\rm SK}=0.5$, although this was found to be somewhat process
dependent.\footnote{The optimal choice of $a_{\text{SK}}$ would also
  change if we were to include $\pi^0$ and $B$-hadron decays in our
  simulation or if we were performing charge-hadron subtraction or
  including a calorimeter simulation~\cite{Cacciari:2014gra}.}
For the case of $pp\rightarrow t\bar{t}$, the dependence of the jet mass on the choice of grid parameter is
shown in \Fig{fig:SK-scan}.
We find that the position of
  the top peak depends significantly on the value of  $a_{\rm SK}$ with small (large) values of $a_{\text{SK}}$ showing a clear
  undersubtraction (oversubtraction).
  In our simulations, $a_{\rm SK}
  \simeq 0.5$ shows only a small average bias, so we take this value as our benchmark.

We now study the effects of pileup on the mass resolution in groomed jets, following the same analysis strategy as \Sec{sec:mass-resol}.
Since the behavior is quite similar to that observed previously, we focus here only on top events, with the $W$ and Higgs cases discussed
in~\App{sec:WH-pileup}.

\begin{figure}[t]
  \centering
    \subfloat[]{%
    \includegraphics[trim={0 0 0 0.5cm},clip,width=0.47\textwidth]{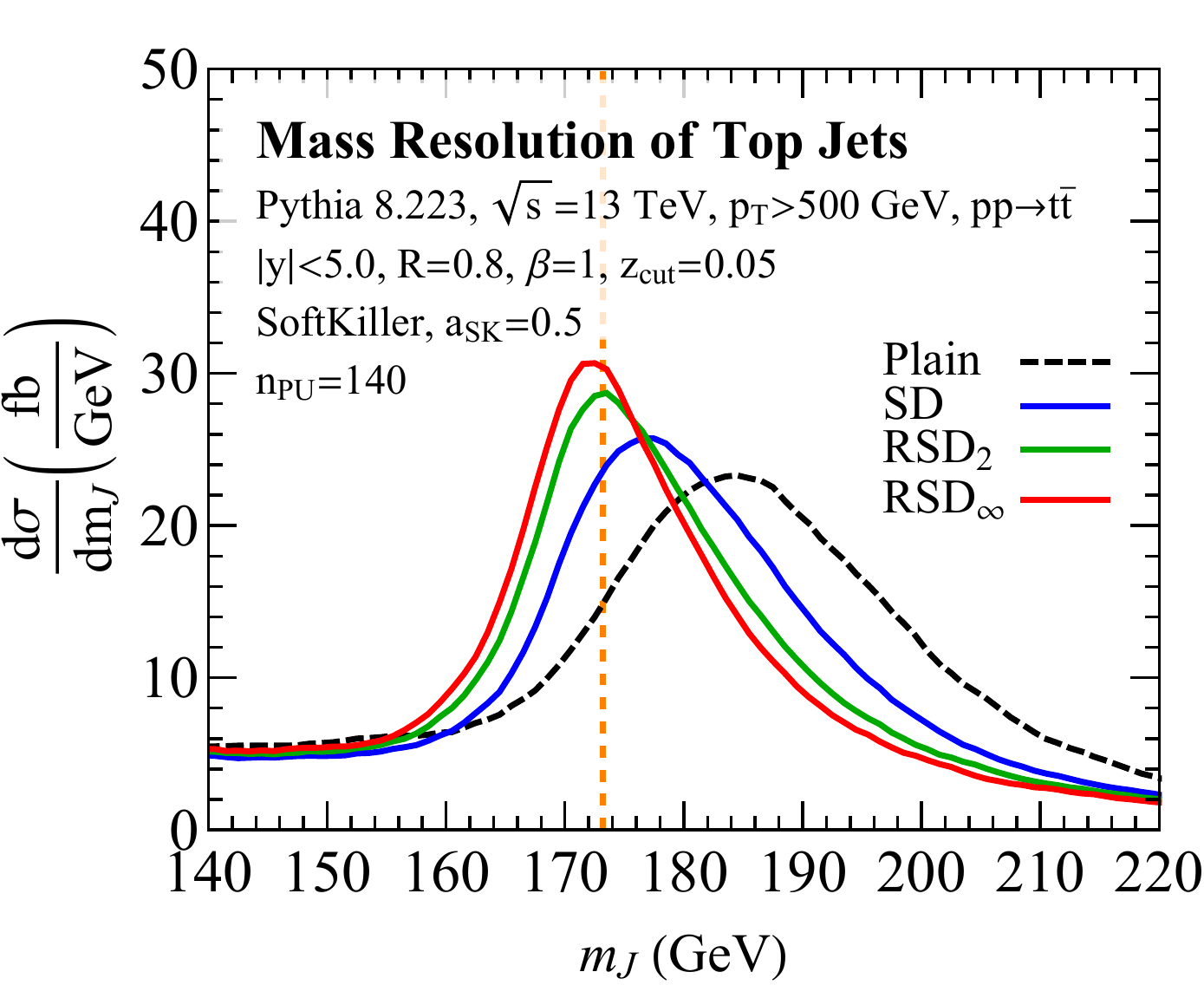}%
    \label{fig:m-PU-top_dist}}%
  \qquad
  \subfloat[]{\label{fig:m-PU-top_shift}%
  \begin{tabular}[b]{c}
    \includegraphics[trim={0 1.52cm 0 0},clip,width=0.4\textwidth]{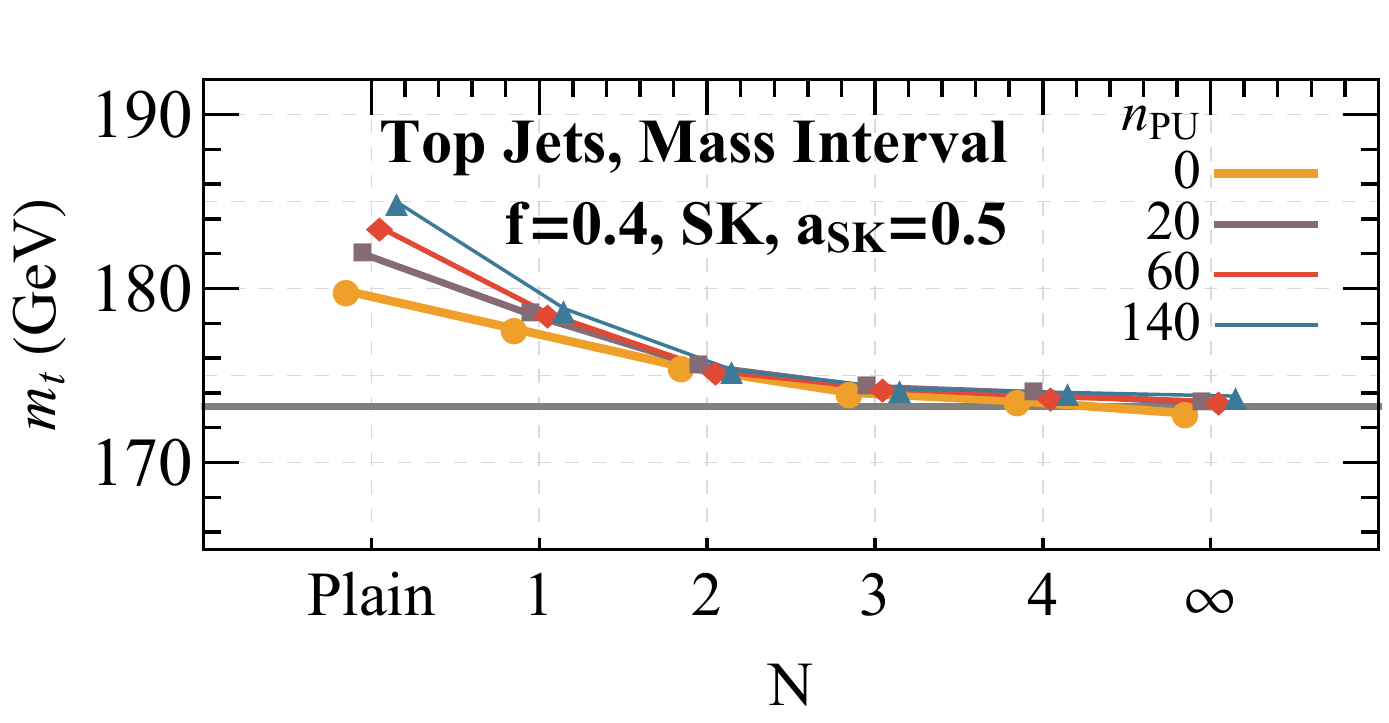} \\[2mm] % 
    \includegraphics[trim={-0.2cm 0.2cm 0.1cm 0.5cm},width=0.39\textwidth]{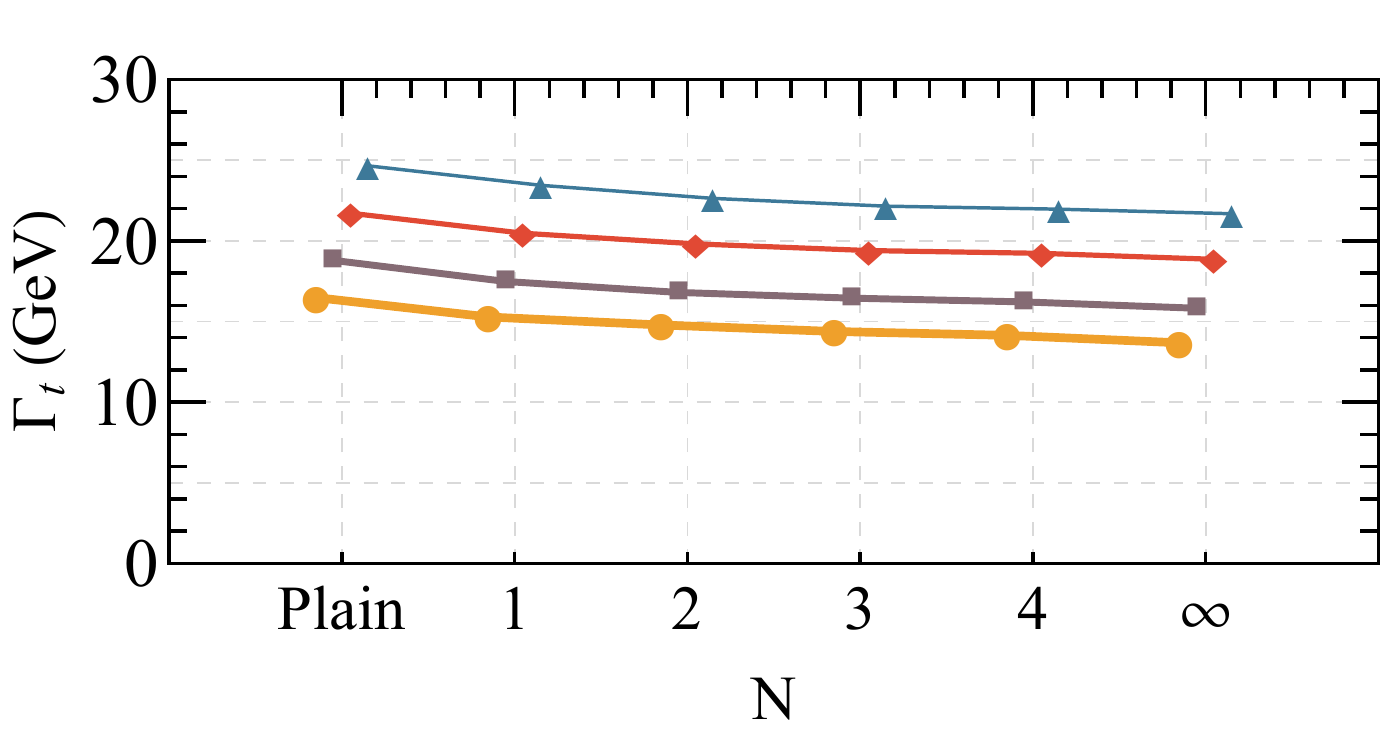}%
    \end{tabular}}
  \caption{(a) Same as \Fig{fig:top_mass_distribution_dist}, but with
    140 pileup vertices and using a SoftKiller parameter of
    $a_{\rm SK}=0.5$.  (b) Central mass values and widths as a
    function of $N$ for several pileup multiplicities, using the
    benchmark RSD/SoftKiller parameters and $f = 0.4$.}
  \label{fig:m-PU-top}
\end{figure}

We use the same $pp \to t \bar{t}$ process as in \Sec{sec:top_mass}, with the \textsc{Pythia} top mass at 173.2 GeV.
In \Fig{fig:m-PU-top_dist}, we show the top jet mass distribution with the addition of 140 pileup vertices.
With SoftKiller but without any jet grooming, there is a 12 GeV shift in the reconstructed top mass peak.
(The shift would be larger than 200 GeV if SoftKiller were not applied.)
The shift decreases to about 5 GeV after applying SD, with additional improvements in going to RSD$_2$ and RSD$_\infty$.
Though the shape of the top mass distribution is not nearly as
symmetric as without pileup, RSD$_\infty$ does restore some of the
symmetry of the top mass distribution compared to SoftKiller
alone.\footnote{As shown in \Ref{Cacciari:2014gra}, one can get better performance for SoftKiller alone by increasing the grid-size parameter $a_{\rm
    SK}$. Even if this works for correcting for the mass
  shift, however, it results in much broader peaks than what we obtain here by
  combining SoftKiller with (R)SD.}

In \Fig{fig:m-PU-top_shift}, we show the central mass value and width after the application of SoftKiller and RSD$_N$ for several pileup levels.
As $\nPU$ increases from 0 to 140, the central top mass value and the width increases monotonically, even with the application of SoftKiller.
By applying more layers of RSD, the central top mass decreases toward
the physical value, with somewhat improved stability across the different pileup levels.
The best performance is obtained for RSD$_\infty$, in agreement with the analysis of \Sec{sec:top_mass}, and the mass difference is less than 5 GeV even at the highest pileup level.
This shows that, as well as improving the resolution of observables
such as the jet mass, grooming with RSD somewhat improves the
stability of the distributions as a function of the number of pileup
vertices $\nPU$.
That said, there are still substantial distortions to the width of the
top mass distribution even after RSD$_\infty$, reflecting the
underlying challenge of pileup at the LHC.

%======================================================================
\subsection{Mass resolution with the area--median method}
\label{sec:PU-area}
%======================================================================

Let us now consider pileup mitigation using RSD in conjunction
with the area--median method~\cite{Cacciari:2007fd,Cacciari:2008gn}.
This removal technique is widely used in experimental analyses, and we
therefore provide a short study of its use with RSD groomed jets.

\begin{figure}[t]
  \centering
    \subfloat[]{%
    \includegraphics[trim={0 0 0 0.5cm},clip,width=0.44\textwidth]{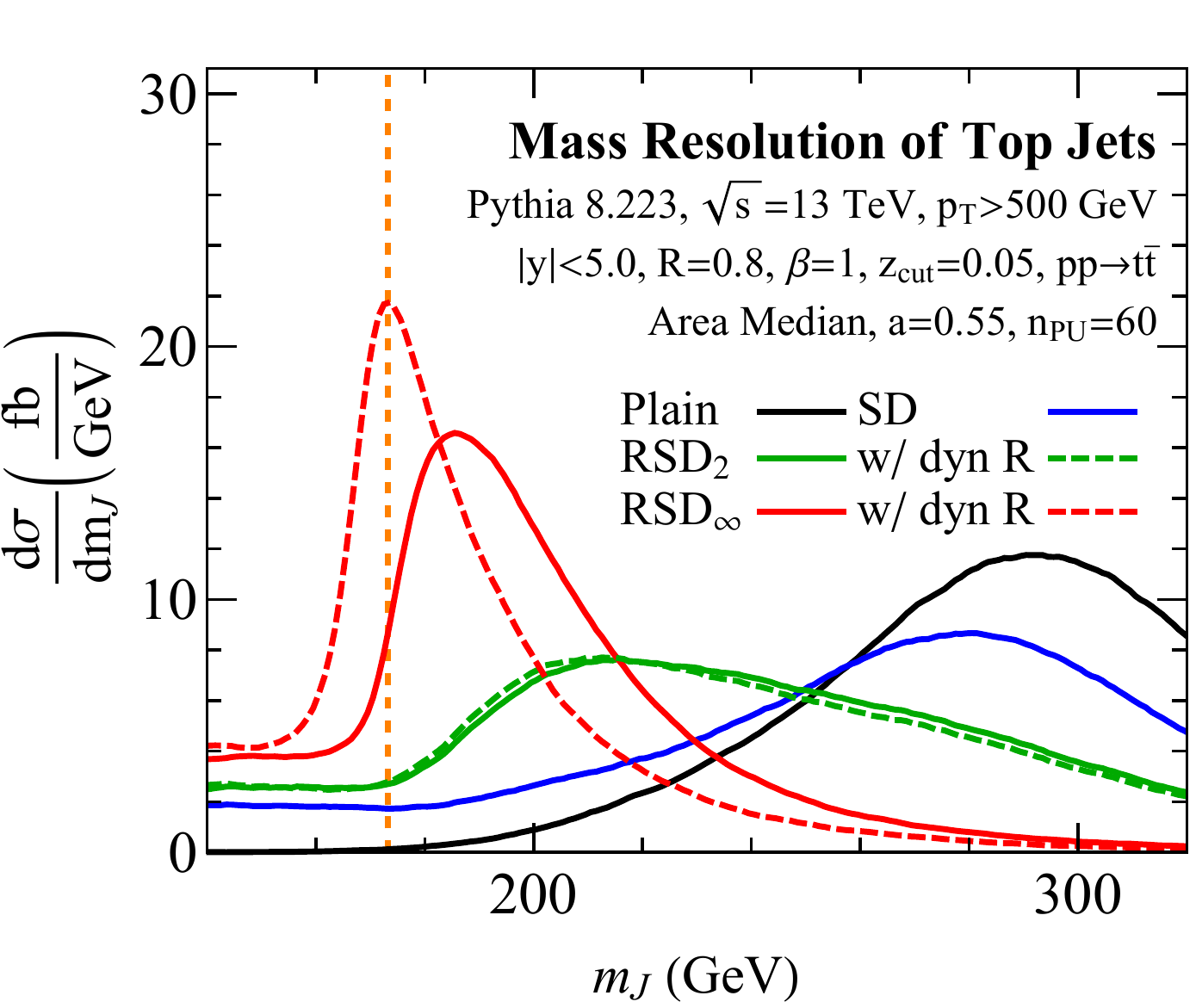}%
    \label{fig:AM-PU-top_dist}}%
  \qquad
  \subfloat[]{\label{fig:AM-PU-top_shift}%
  \begin{tabular}[b]{c}
    \includegraphics[trim={0 1.52cm 0 0},clip,width=0.4\textwidth]{{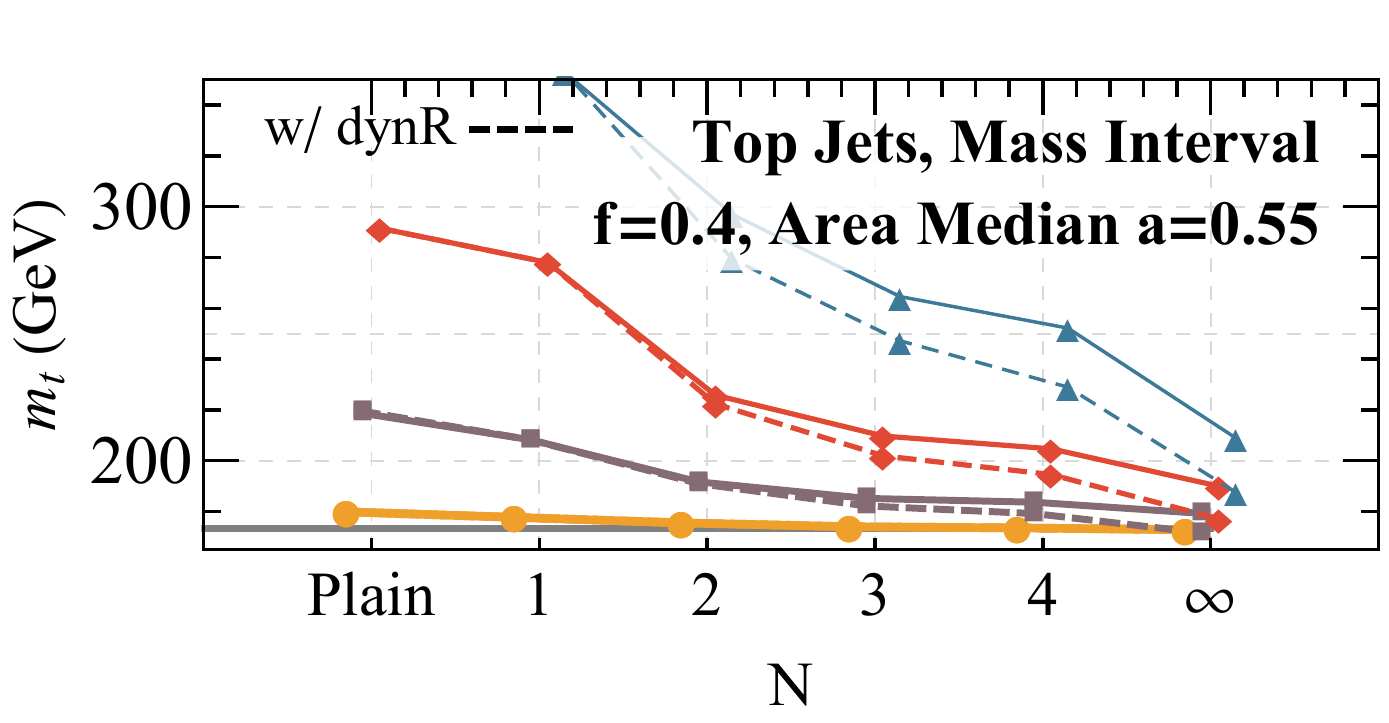}} \\[2mm] % 
    \includegraphics[trim={-0.2cm 0.2cm 0.1cm 0.5cm},width=0.39\textwidth]{{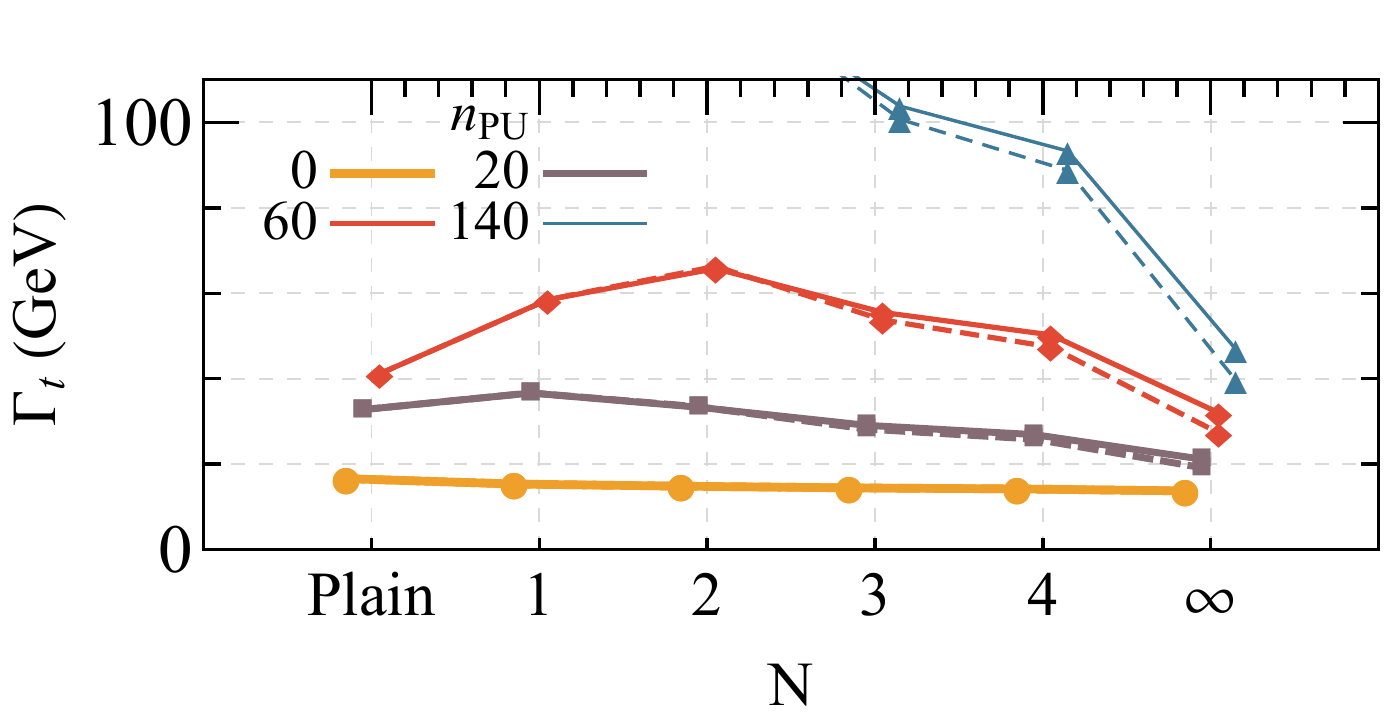}}%
    \end{tabular}}
  \caption{(a)  Similar to \Fig{fig:m-PU-top_dist}, but with 60 pileup vertices and using an area--median parameter of $a=0.55$.  Note the larger mass range shown and the inclusion of both fixed-$R_0$ and dynamic-$R_0$ results.  (b) Same as \Fig{fig:m-PU-top_shift} but for the area--median method applied to RSD with dynamic $R_0$.}
  \label{fig:area-median}
\end{figure}

Generally speaking, combining the area--median procedure with grooming
requires more than just subtracting the jet after grooming.
Indeed, since the grooming procedure imposes kinematic constraints on
subjets, one wants to apply the subtraction procedure directly on the
subjets, so that these kinematic constraints use quantities which have
been corrected for pileup (see e.g.~\cite{Altheimer:2013yza}).
In the case of RSD (and mMDT/SD), this means that, at each step
of the declustering procedure, we should apply an area--median
subtraction to both subjets before imposing the SD condition in
\Eq{eq:SD-crit}.\footnote{A similar philosophy is also recommended, but sadly not always implemented, when using other grooming techniques like pruning and trimming.}
Note that even in the case of RSD$_\infty$ where the final groomed jet
has zero area, intermediate subjets have a non-zero area, so
subtracting the intermediate subjets is crucial.

The resulting mass peak is shown in~\Fig{fig:AM-PU-top_dist}, for
both a dynamic and fixed $R_0$ implementation of RSD for 60 pileup
events, using the area--median parameter $a=0.55$.
We find that only the
most aggressive algorithm RSD$_\infty$, preferably using dynamic
$R_0$ from \Sec{sec:dynamicR}, succeeds at reconstructing
the top peak.
This is confirmed by \Fig{fig:AM-PU-top_shift}, where we
show the central mass values and widths for various pileup levels.
Despite the fact that RSD$_\infty$ with dynamic $R_0$ does recover the top mass peak, one
should still note that the mass resolution (and median peak position)
significantly worsens with increasing pileup multiplicity, encouraging
the investigation of more recent particle-level pileup mitigation techniques for
future runs of the LHC.

%======================================================================
\section{Bottom-Up Soft Drop for event-wide grooming}
\label{sec:BUSD}
% ======================================================================

We have seen that the default RSD algorithm in \Sec{sec:recursive-SD} yields sensible
grooming behaviors across a wide range of applications, with excellent overall performance for $N = \infty$.
It is possible, however, to obtain similar results with a different
approach.
In this section, we introduce an alternative to RSD$_\infty$ called Bottom-Up Soft Drop (BUSD), which is also available in {\tt{RecursiveTools}}~($\ge$2.0.0) through \texttt{fastjet-contrib} \cite{fjcontrib}.

The default RSD is a top-down algorithm, where the SD condition in
\Eq{eq:SD-crit} is imposed by declustering a jet starting from its
clustering tree.
By contrast, BUSD imposes \Eq{eq:SD-crit} as part of a (re)clustering
procedure, effectively starting from the leaves of the clustering
tree.
BUSD can either be applied to a single jet (\emph{local BUSD}) or to
the event as a whole (\emph{global BUSD}).

With either BUSD approach, the SD criteria is applied at each pairwise combination
during the reclustering stage, somewhat like
pruning~\cite{Ellis:2009me}.
In the \texttt{fastjet-contrib} implementation, this is achieved through a modified recombination scheme, such that at each step of the reclustering with a large-$R$ C/A algorithm, the pseudojet obtained from combining particles $i$ and $j$ with smallest distance $d_{ij} = \Delta R_{ij}/R_0$ is given by
\begin{equation}
  p_{ij} =
\left\{
	\begin{array}{ll}
		\max(p_i,p_j)  & \mbox{if \Eq{eq:SD-crit} fails,}  \\
		p_i + p_j & \mbox{otherwise,}
	\end{array}
\right.
\end{equation}
where the maximum is defined by $p_t$.
In the definition of $d_{ij}$, we choose $R_0$ to match the SD criteria in \Eq{eq:SD-crit}.
For the studies below, we always match the parameter $R_0$ to the jet radius $R$.

With local BUSD, the reclustering is applied to an individual jet found by another jet algorithm.
While one cannot obtain finite $N$ results with BUSD, the bottom-up algorithm provides results that are very similar to the $N=\infty$
top-down approach of $\RSDinf$.
This is expected, since local BUSD uses the same initial jet constituents as $\RSDinf$.

With global BUSD, the full event is clustered into a single large C/A tree.
This provides an event-wide grooming strategy that does not require any specific jet definition.
After grooming with global BUSD, the groomed event contains only a subset of the initial particles, so any jet definition can be used to cluster the remaining particles.
The resulting jets are guaranteed to have zero active area, without any additional treatment required for each individual jet.

\begin{figure}
  \centering
  \subfloat[]{%
    \begin{minipage}{0.45\linewidth}
      \includegraphics[width=1.0\textwidth,trim={0cm 1.57cm 0 0cm},clip]{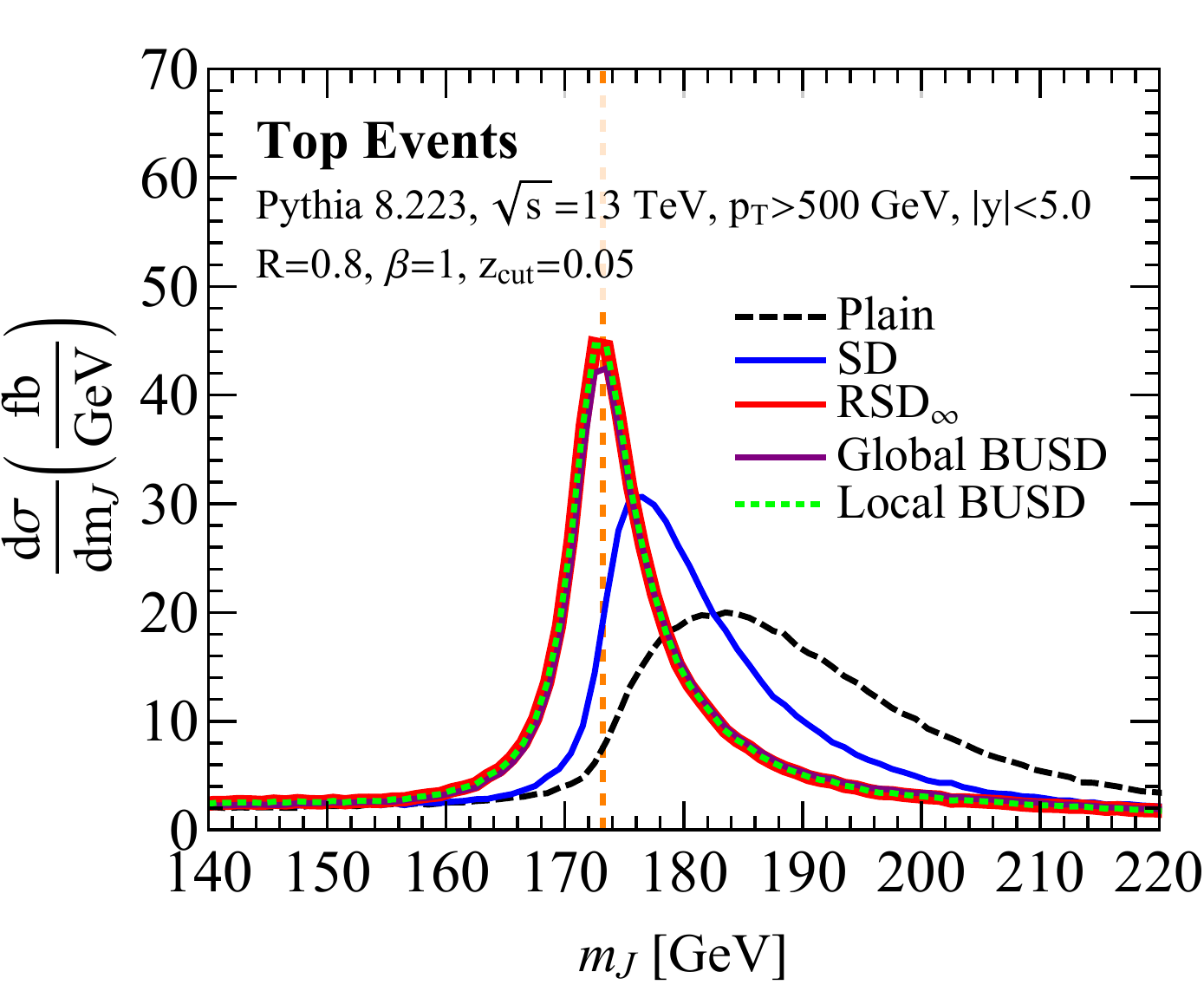}      
      \includegraphics[width=1.0\textwidth,trim={-1.3cm 0 0 0cm}]{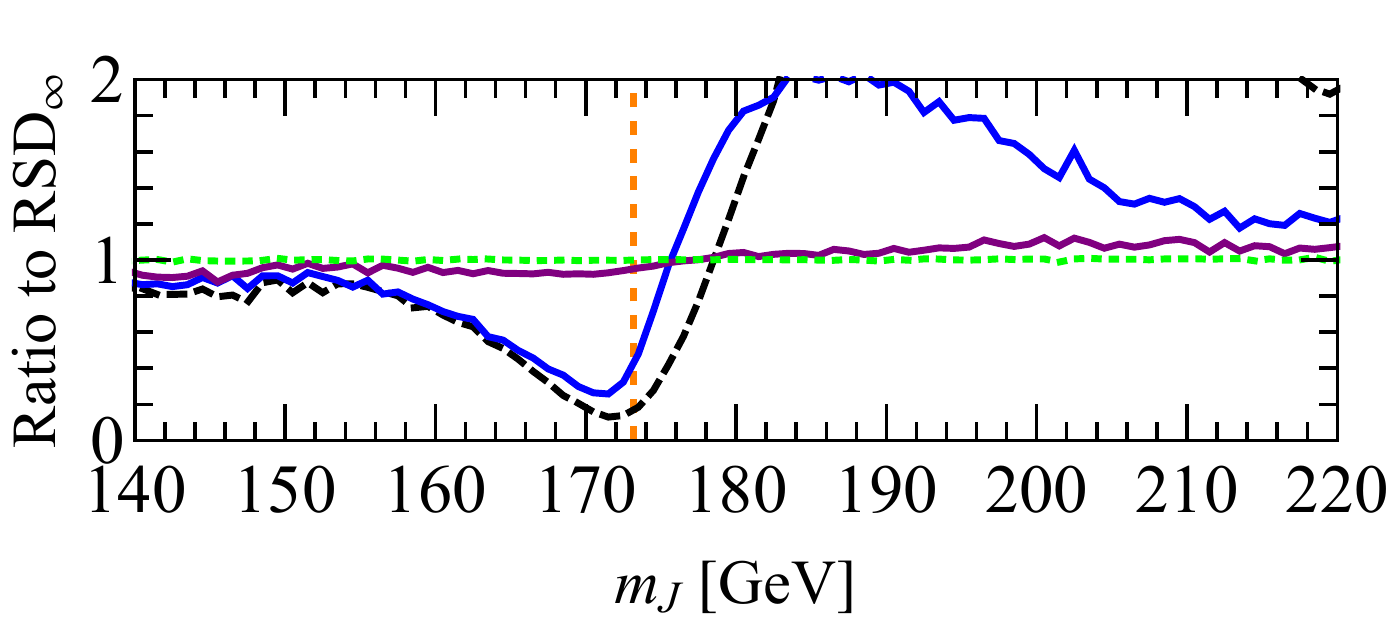}
    \end{minipage}%
    \label{fig:global-grooming-noPU}}%
  \qquad
  \subfloat[]{%
    \begin{minipage}{0.45\linewidth}
      \includegraphics[width=1.0\textwidth,trim={0cm 1.57cm 0 0cm},clip]{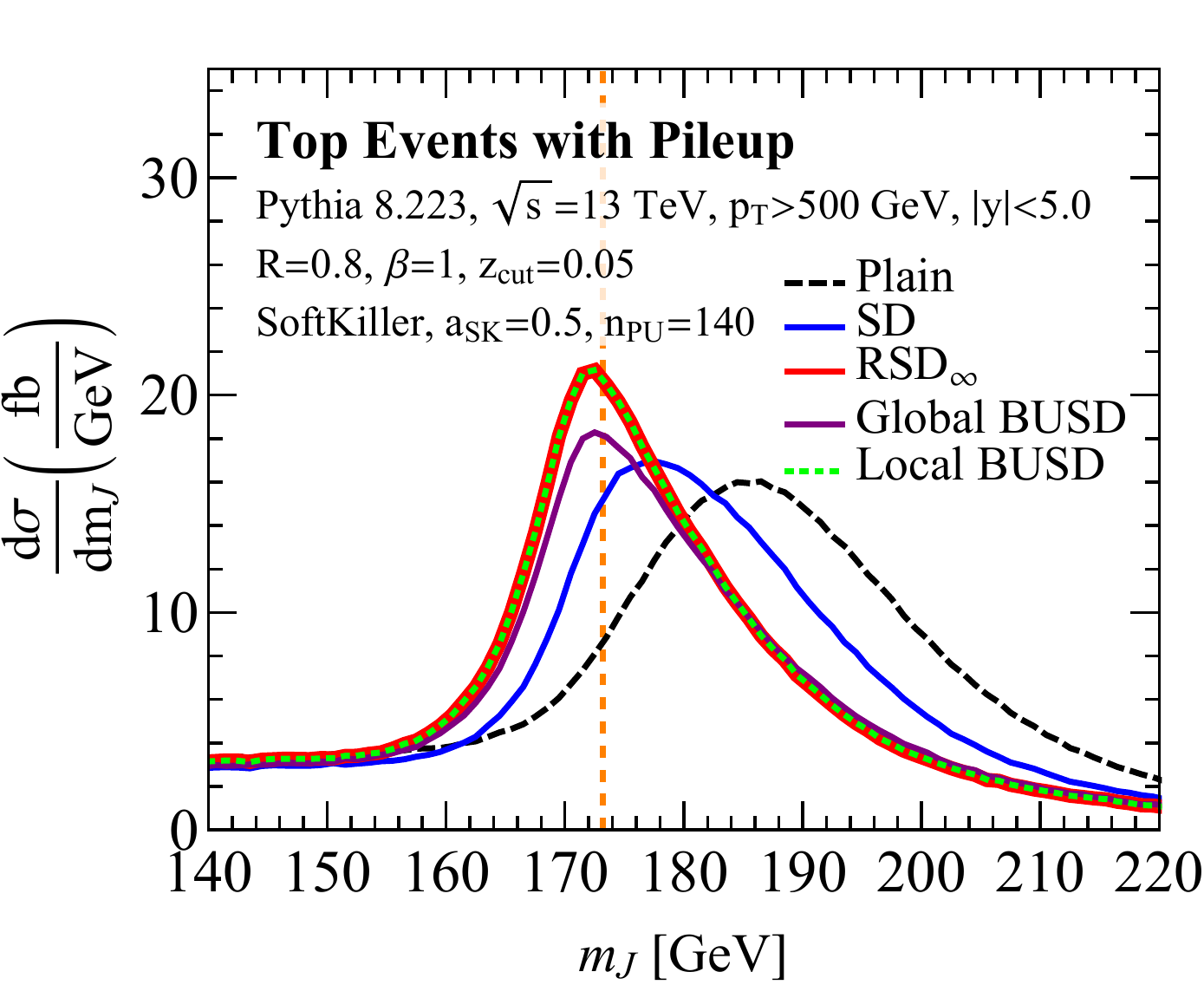}
      
      \includegraphics[width=1.0\textwidth,trim={-1.3cm 0 0 0cm}]{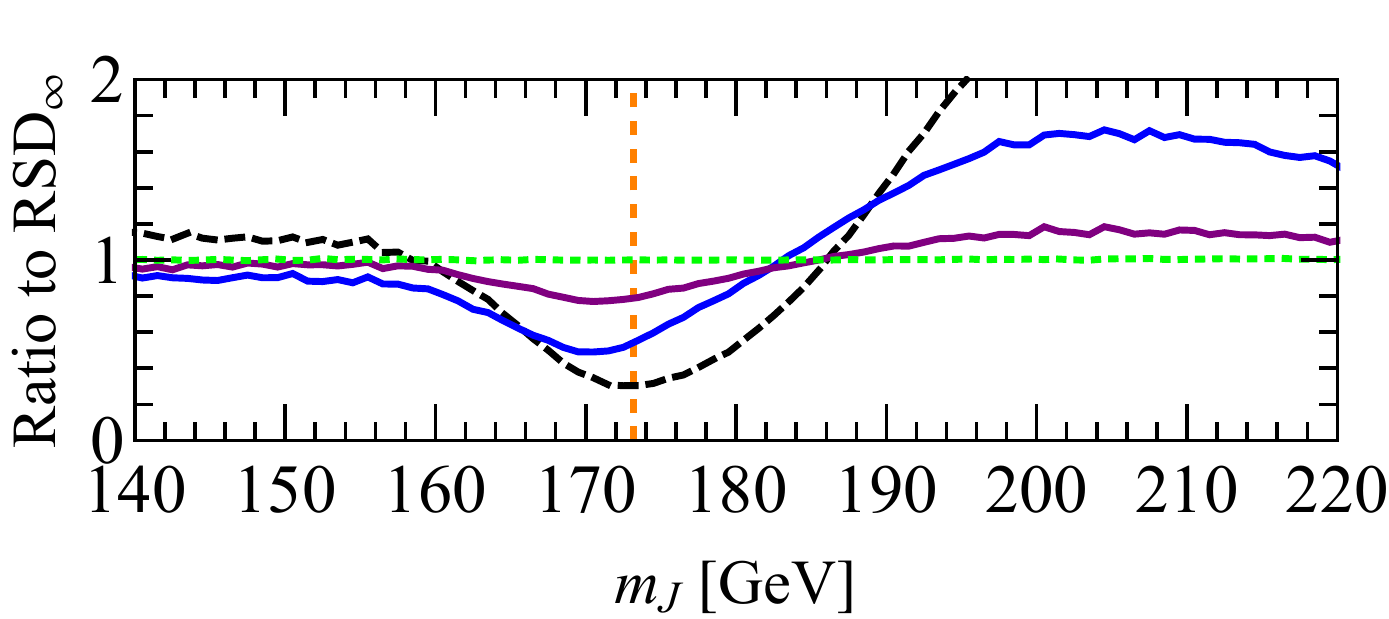}
    \end{minipage}%
    \label{fig:global-grooming-PU}}
  \caption{  Behavior of global and local BUSD on $t\bar{t}$ events
    with the default RSD parameters.  (a)  Performance without pileup,
    to be compared to \Fig{fig:top_mass_distribution_dist}.  (b)
    Performance with 140 pileup vertices plus SoftKiller, to be compared to
    \Fig{fig:m-PU-top_dist}.  In both panels, the ratios to
    RSD$_{\infty}$ are shown at the bottom.}
  \label{fig:global-grooming}
\end{figure}

The behavior of BUSD is shown in \Fig{fig:global-grooming-noPU} for the same $pp \to t \bar{t}$ sample from \Sec{sec:top_mass}.
Without pileup, $\RSDinf$, local BUSD, and global BUSD give nearly identical results on the top jet mass distribution.
This suggests that globally grooming an event before the jet clustering stage could be a practical alternative to standard jet-based grooming.

We test the robustness of BUSD to pileup in \Fig{fig:global-grooming-PU}.
As in \Sec{sec:mass_PU}, we overlay 140 pileup vertices and apply the SoftKiller algorithm with grid parameter $a_{\rm SK}=0.5$.
There is still nearly identical behavior for $\RSDinf$ and local BUSD, with somewhat degraded performance seen using global BUSD.
So while global BUSD grooming still performs better than the original SD in reconstructing the top mass, it is outperformed by $\RSDinf$ (and local BUSD) in
extreme environments.

%======================================================================
\section{Conclusions}
\label{sec:conclusions}
%======================================================================

Recursive Soft Drop (RSD) is a generalization of the original mMDT/SD
algorithm, which has already seen successful applications at the LHC.
By recursively applying the SD declustering procedure $N$ times,
RSD$_N$ can remove jet contamination more efficiently and with looser
SD parameters than original SD.
With benchmark parameters $\beta=1$ and $\zcut=0.05$, we found that RSD grooming performs particularly well in the fully-recursive $N=\infty$ limit.
This limit also shows an improved robustness against pileup effects, a
feature which is likely connected to the fact that RSD$_\infty$ yields
groomed jets with formally zero area.

Jet grooming with RSD provides several refinements over previous techniques, notably by increasing the robustness to
non-perturbative effects and by improving the mass resolution for boosted heavy resonances such as $W$ bosons, top quarks, and Higgs bosons.
In the context of jet tagging, the additional declustering layers have
a modest impact on QCD backgrounds, so much of the gains in tagging
performance come from the improved jet mass resolution.
While RSD$_{N-1}$ seems to be the natural choice for grooming $N$-prong objects, we have
noticed that adding more grooming layers, e.g.\ using RSD$_N$, comes with further resolution gains.
In the end, we recommend the use of RSD$_\infty$, since it offers the same or better performance in our case studies with no discernible downsides.
The one possible
exception is for boosted $W$ bosons, where RSD$_\infty$ gave a better
$W$ boson resolution compared to SD but at the expense of shifting the $W$ peak
location by ${\cal{O}}(1)$~GeV.
That said, this shift could be minimized by using lower values of $\zcut$ or higher values of $\beta$.

For pileup mitigation, RSD works best when paired with a particle-level removal algorithm such as SoftKiller.
In the presence of pileup, we found substantial improvements in the groomed mass resolution after grooming with RSD when compared
to the original SD procedure.
We recommend the use of RSD$_\infty$ in high luminosity environments,
since this always performed better than RSD with finite $N$.
It is also possible to use RSD with area-based pileup subtraction methods, as long as the corrections are applied on the finite-area subjets that enter into the SD condition.

An interesting alternative to RSD$_\infty$ is Bottom-Up Soft Drop
(BUSD).
In its local variant, i.e.\ applied to a single jet, it shows
similar performance to RSD$_\infty$, with comparable resilience
to pileup effects.
BUSD also admits a global implementation, where it
is applied at the event-wide level to groom an event without
committing to a particular jet algorithm.
This latter variant shows,
however, a slightly larger sensitivity to pileup than RSD$_\infty$.

Since RSD is closely related to the mMDT/SD algorithms, it shares the advantage of retaining analytic tractability.
We look forward to future studies aimed at high-precision and systematically-improvable analytical results of tagging rates.
These could be achieved with existing frameworks to
leading-logarithmic accuracy, and could potentially be extended to
higher logarithmic accuracy through a suitable extension of the
CAESAR~\cite{Banfi:2004yd} and ARES~\cite{Banfi:2014sua,Banfi:2016zlc}
methods or through a modified factorization theorem~\cite{Frye:2016okc} in
soft-collinear effective theory~\cite{Bauer:2000yr,Bauer:2001ct,Bauer:2001yt,Bauer:2002nz}.
An interesting potential application of RSD is in defining the short-distance top quark mass using light grooming \cite{Hoang:2017kmk}. 

Finally, RSD could also find useful applications in the context of heavy ion collisions, where jet grooming can provide a powerful probe
into the effects of the medium on the momentum sharing $z_g$~\cite{Sirunyan:2017bsd,Chien:2016led,Mehtar-Tani:2016aco,Milhano:2017nzm}.
By adjusting the number of grooming layers $N$, one can achieve a more aggressive grooming while retaining more of the underlying hard scattering process.
Furthermore, the groomed energy fractions $z_{g,i}$ obtained at every SD layer $i$ may provide an additional handle to study the propagation of
partons through the quark-gluon plasma at multiple resolution scales.

%======================================================================
%======================================================================

\section*{Acknowledgements}
We would like to thank Yang-Ting Chien, Andrew Larkoski, and Ian Moult
for useful discussions, as well as Mrinal Dasgupta and Simone Marzani
for comments on the manuscript.
FD is supported by the SNF grant P2SKP2\_165039.
FD and JT are supported by the Office of High Energy Physics of the U.S. Department of Energy (DOE) under grant DE-SC-0012567.
GS's work is supported in part by the French Agence Nationale de la
Recherche, under grant ANR-15-CE31-0016.
LN's work is supported by the DOE under Award Number DE-SC-0011632, and the Sherman Fairchild  fellowship.

%======================================================================
%======================================================================

\appendix

%======================================================================
%======================================================================

%======================================================================
\section{Behavior at fixed order} \label{sec:fixed_order}
%======================================================================

In this appendix, we study the behavior of RSD at fixed order in $\as$.
To this end, we use \texttt{Event2}~\cite{Catani:1996jh,Catani:1996vz}
to produce a sample of boosted heavy bosons decaying (at tree level)
to a quark-antiquark pair.
We start from an $e^+e^-$ collision at a given center-of-mass energy
$M\equiv\sqrt{s}$, corresponding to the mass of the heavy boson.
Then, we boost the partonic system transversely such that the heavy boson is traveling along the $x$ axis of the collision with transverse momentum $p_t$.
In what follows, we fix $M=100$~GeV and $p_t=500$~GeV.
We cluster the event with the anti-$k_t$ algorithm with
$R=0.8$, and we apply RSD$_N$ with $\beta=1$, $z_{\text{cut}}=0.2$,
varying the number of layers $N$.
Note the use of a larger $z_{\text{cut}}$ value compared to our recommended default to make the impact of RSD more visible.

The motivation for studying such boosted systems is that it provides a
source of jets with 2, 3, and 4 partons at respective orders
$\alpha_s^0$, $\alpha_s^1$, and $\alpha_s^2$, allowing us to study
the effects of several layers of RSD.

\begin{figure}[t]
  \centering
  \subfloat[]{%
    \begin{minipage}{0.45\linewidth}
    \includegraphics[width=1.0\textwidth]{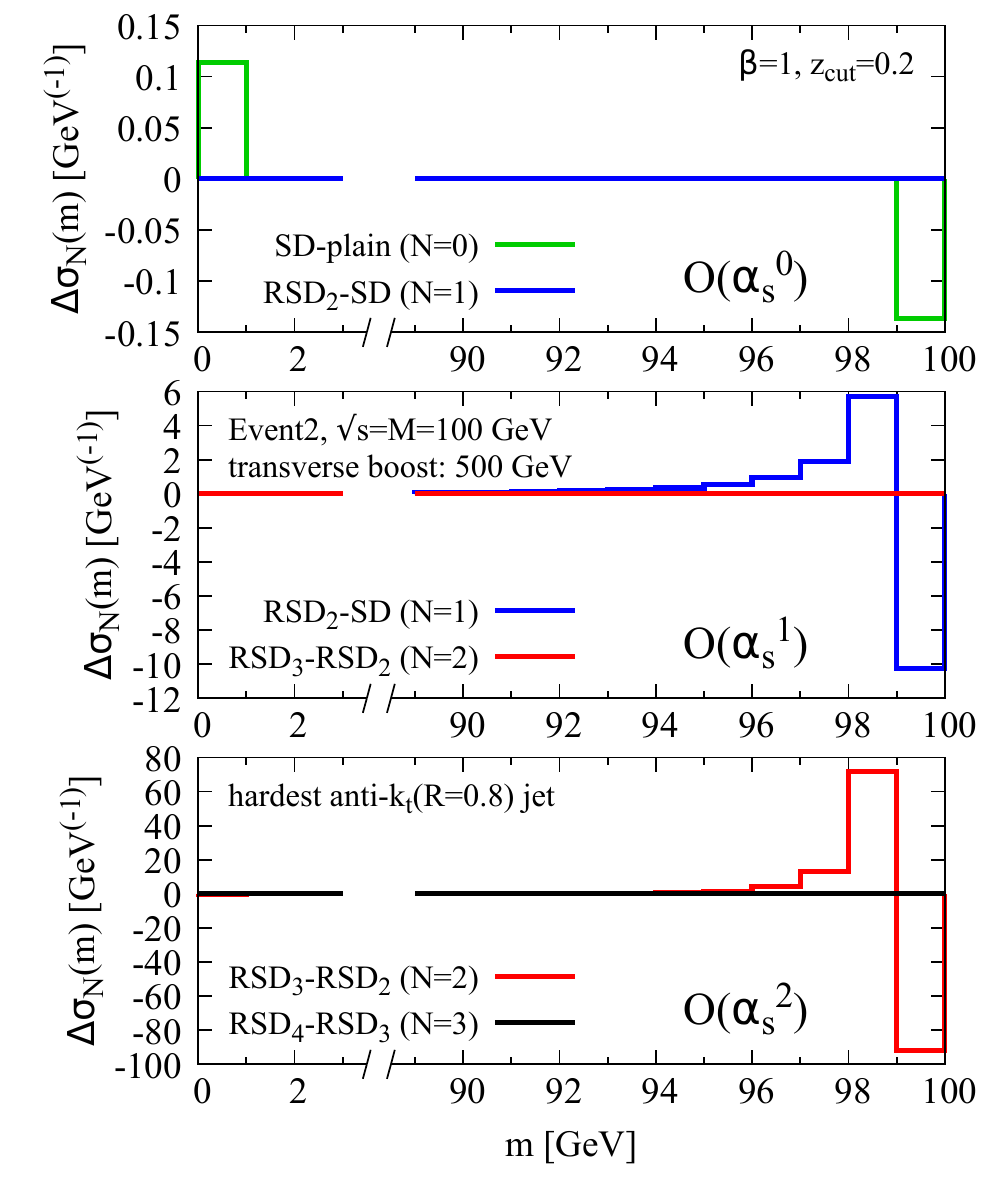}   
    \end{minipage}%
    \label{fig:event2-mjet}}%
  \qquad
  \subfloat[]{%
    \begin{minipage}{0.45\linewidth}
    \includegraphics[width=1.0\textwidth]{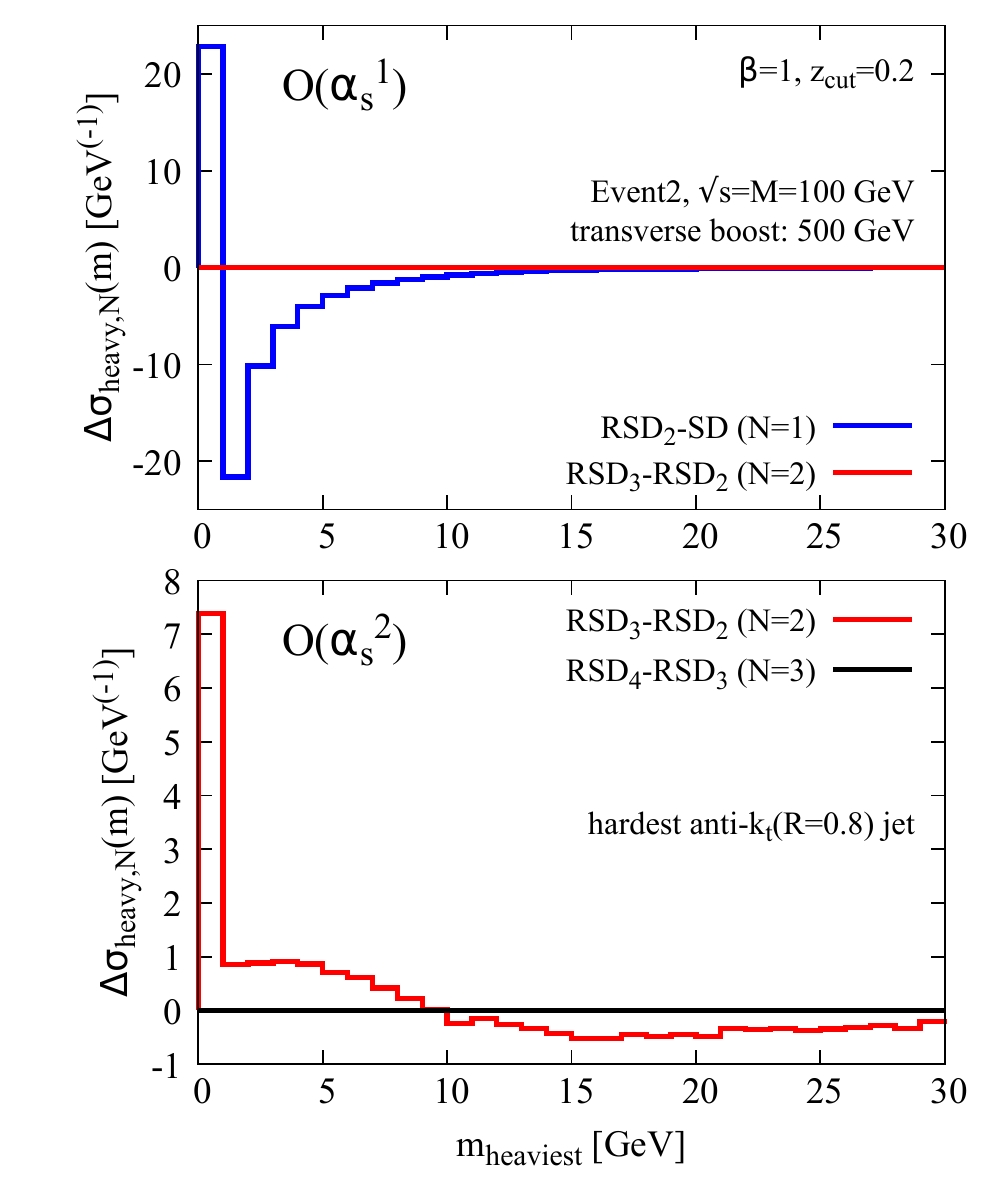}  
    \end{minipage}%
    \label{fig:event2-mhard}}
  \caption{Fixed-order coefficients from \texttt{Event2} for (a) the
    jet mass distribution and (b) the heavier subjet mass
    distribution. The plot shows \Eq{eq:massdifference}, the
    difference in the jet mass spectrum obtained with consecutive
    layers of RSD.}
  \label{fig:event2}
\end{figure}

Let us start by discussing the jet mass in \Fig{fig:event2-mjet}. 
Instead of plotting the mass distribution after applying RSD$_N$, we
plot the difference
\begin{equation}
\label{eq:massdifference}
  \Delta \sigma_N(m) = \frac{1}{\sigma_{\text{Born}}}\left[
    \left.\frac{d\sigma}{dm}\right|_{\text{RSD}_{N+1}} -
    \left.\frac{d\sigma}{dm}\right|_{\text{RSD}_N} \right],
\end{equation}
i.e.\ the difference in the mass spectrum between two consecutive
layers of RSD.
From top to bottom, \Fig{fig:event2-mjet} shows $\Delta\sigma_N$ for tree-level
events ($e^+e^-\to q\bar q$) at ${\cal{O}}(\alpha_s^0)$, the order $\alpha_s$ correction, and the order $\alpha_s^2$ correction.

At ${\cal{O}}(\alpha_s^0)$, there is either one or two partons in the
jet.
For one parton, the jet mass is always zero, irrespectively of
the number of layers of RSD.
Jets with two partons, however, have a (plain) jet mass $M$.
Therefore, if we apply SD = RSD$_1$ to such a jet, we can either have a jet with two partons and
a mass $M$, or a jet with one parton and a zero mass, depending on
whether or not the SD condition is satisified.
Since there is no additional substructure to probe in the jet,
applying additional SD layers with RSD$_{N>1}$
will not have any effect.
This expectation is indeed observed in the
top panel of \Fig{fig:event2-mjet}: applying the first layer of
SD yields a decrease of the cross section for $m=M=100$~GeV
and an increase at zero mass, i.e.\ $\Delta\sigma_0\neq 0$, and
additional SD layers have no effect, i.e.\ $\Delta\sigma_{(\ge)1}= 0$.

At order $\alpha_s$, we now have jets with
three partons, which is enough to see a non-trivial effect from RSD$_2$.
Indeed, if we have a jet with three partons for which the first declustering passes the SD
condition, the second SD layer will sometimes be applied on
a subjet itself made of two partons; if this system fails the SD condition, the RSD$_2$ mass will be smaller than the RSD$_1$
mass, with all subsequent layers being ineffective.
This is seen in the middle panel of \Fig{fig:event2-mjet}, with $\Delta\sigma_1$
showing a shift towards smaller masses and $\Delta\sigma_{(\ge)2}= 0$.

Unsurprisingly, the same pattern is observed at order $\alpha_s^2$, just one layer further down,
with $\Delta\sigma_2$ showing a decrease in mass while $\Delta\sigma_{{\ge}3}=0$.
Note that although the two-loop virtual corrections, contributing at
${\cal{O}}(\alpha_s^2)$, are not available in \texttt{Event2}, they
correspond to events where the jets can have at most two partons and
so do not contribute to $\Delta\sigma_{N\ge 1}$.
Ultimately, we see that each successive SD layer further grooms the jet.
For the $N^{\text{th}}$ layer of SD to be effective, one needs
at least $N+1$ partons in the jet, hence $\Delta\sigma_N$ being
non-zero starting at order $\alpha_s^N$. 

In addition to the jet mass distribution, we can also study the mass
spectrum of the heavier subjet in a jet, shown in \Fig{fig:event2-mhard}.
This is defined by applying RSD to the jet and taking, in the resulting groomed jet, the heavier of the
two subjets corresponding to the \emph{first} declustering that has passed
the SD condition.
Taking the example of
  \Fig{fig:RSD}, this is the heavier of the RSD
  subjets tagged at vertex ``1'', i.e.\ the heavier of the subjet made
  of the two upper kept particles and the subjet made of the two lower
  kept particles.

At ${\cal O}(\alpha_s^0)$, there can be at most one particle in a
subjet so the heavier subjet mass is always 0.
At higher orders, though, one can get more
partons and hence a non-zero subjet mass.
We study the difference $\Delta \sigma_{\text{heavy},N}(m)$  in \Fig{fig:event2-mhard},
defined analogously to \Eq{eq:massdifference} as the difference in the heavier subjet mass distribution
between $N$ and $N+1$ layers of RSD.
As with the jet mass, we need at least $N+2$ partons in the
jet  to get a non-zero $\Delta\sigma_{\text{heavy},N}$, a situation which starts at order $\alpha_s^N$.
This is confirmed by our \texttt{Event2} simulations where, for
example, $\Delta\sigma_{\text{heavy},1}$ is already different from zero
at ${\cal O}(\alpha_s)$ while $\Delta\sigma_{\text{heavy},2}$ starts
being non-zero at ${\cal O}(\alpha_s^2)$.

A calculation of the logarithmically-enhanced terms appearing in these
distributions could serve as a base for an analytic study of RSD.

% ======================================================================
\section{Additional pileup studies}
\label{app:additionalpileup}

In this appendix, we present additional results for RSD in high pileup conditions.

\subsection{$W$ and Higgs mass resolution with SoftKiller}
\label{sec:WH-pileup}

Analogously to the case of 3-pronged top decays in \Sec{sec:mass_PU}, here we show results for 2-prong boosted
$W$ bosons and 4-prong boosted $H$ bosons.
In all cases, we use the default RSD parameters $\beta = 1$ and $\zcut =0.05$, and apply SoftKiller with $a_{\rm SK} = 0.5$.

\begin{figure}[t]
  \centering
    \subfloat[]{%
    \includegraphics[trim={0 0 0 0.5cm},clip,width=0.46\textwidth]{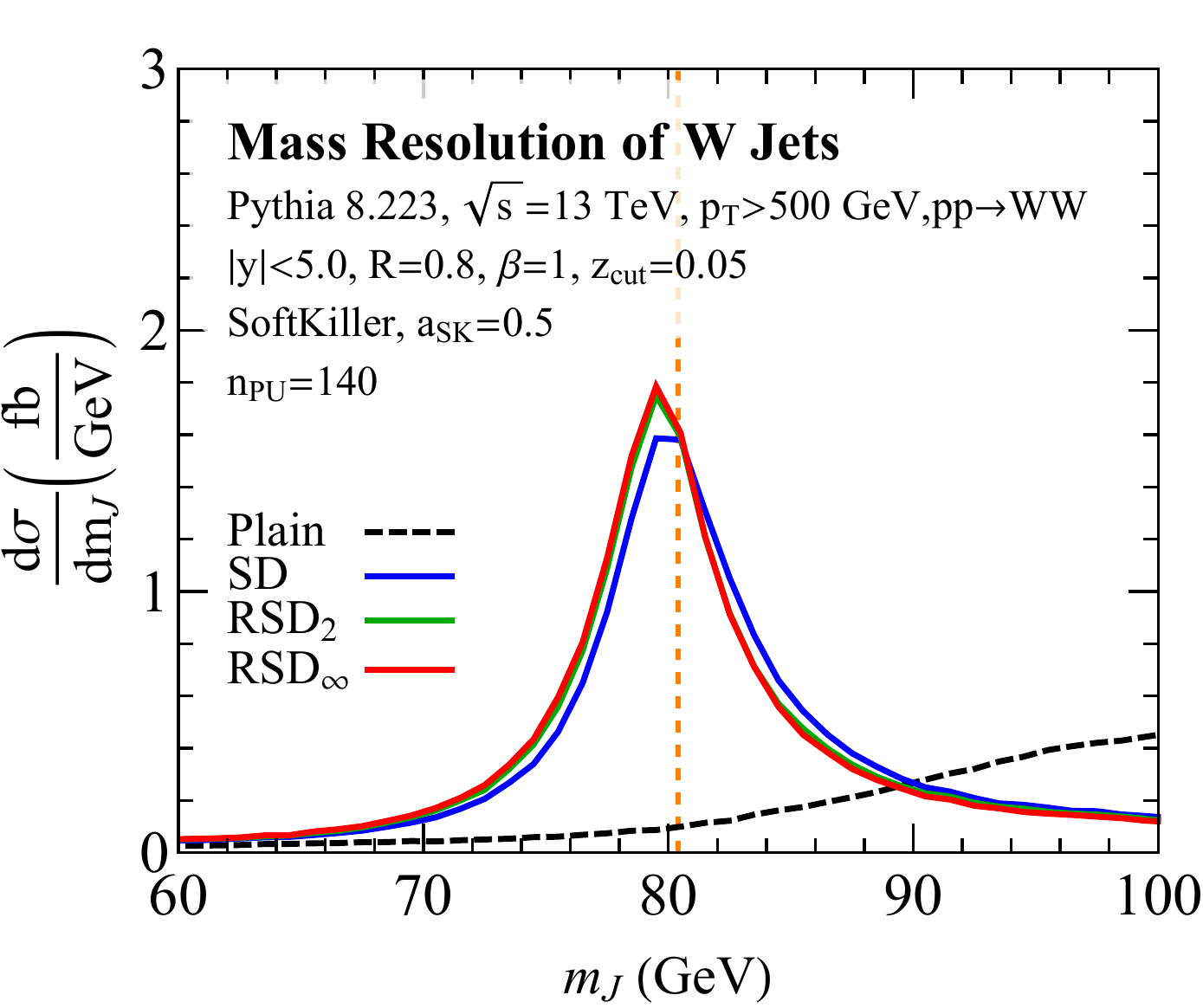}%
    \label{fig:m-PU-WW_dist}}%
  \qquad
  \subfloat[]{\label{fig:m-PU-WW_shift}%
  \begin{tabular}[b]{c}
    \includegraphics[trim={0 1.52cm 0 0},clip,width=0.4\textwidth]{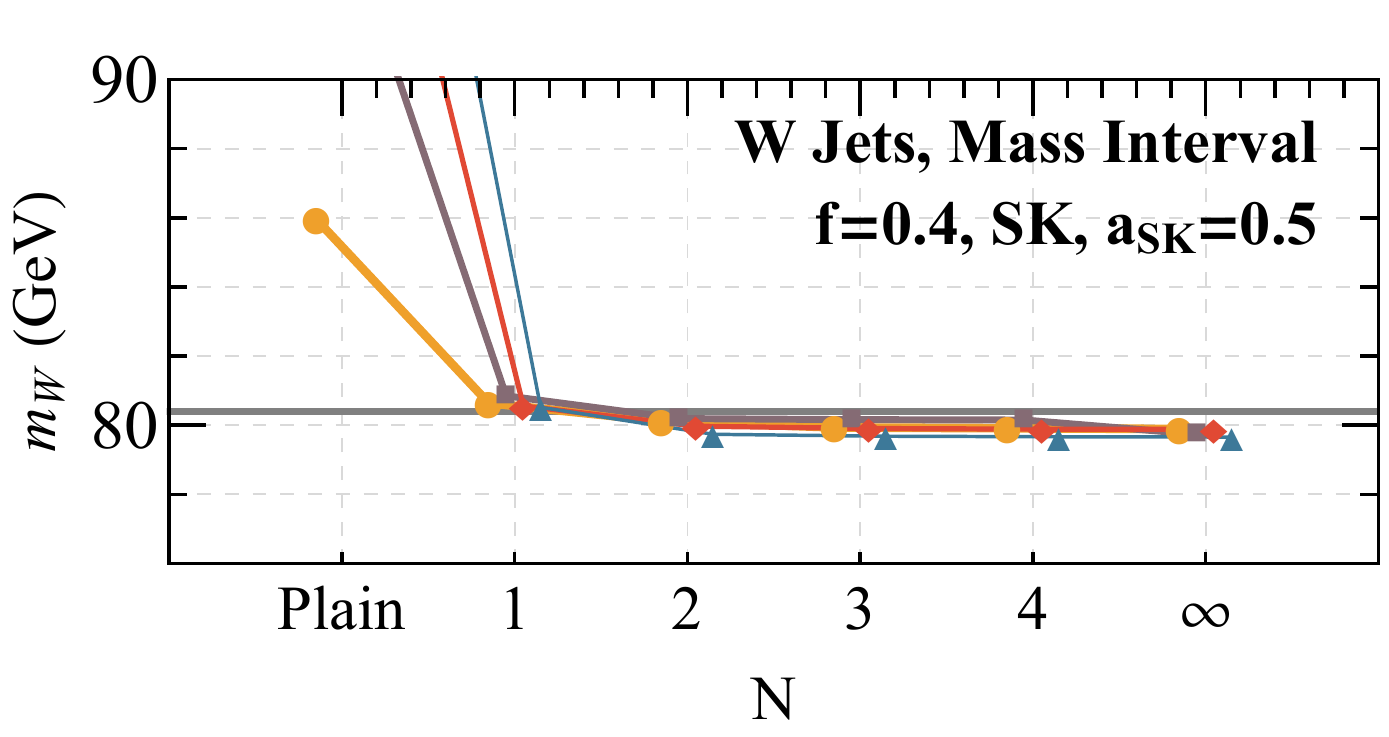} \\[2mm] % 
    \includegraphics[trim={0 0.2cm 0 0.5cm},width=0.39\textwidth]{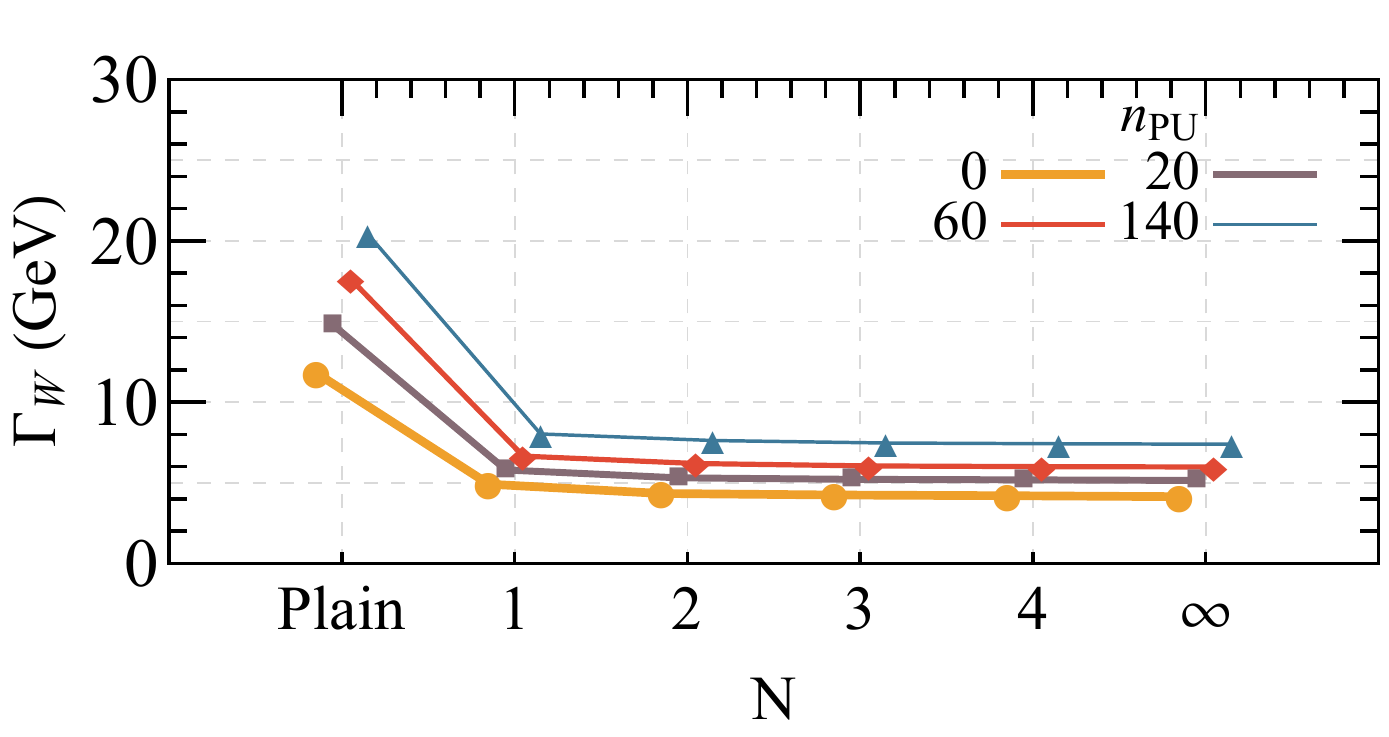}%
    \end{tabular}}
  \caption{Same as \Fig{fig:m-PU-top} but for the $W$ boson sample.}
  \label{fig:m-PU-WW}
\end{figure}

The $W$ mass distribution is shown in \Fig{fig:m-PU-WW_dist}, with the addition of 140 pileup vertices.
This is using the same event samples as in \Sec{sec:W_mass} and can be compared to \Fig{fig:top_mass_distribution_dist}.
Already, the performance is very good for SD alone, though the mass resolution is improved somewhat going to RSD$_2$ or RSD$_\infty$.
Note that the distribution becomes a bit more symmetric, though not nearly as much as in the case without pileup.
In \Fig{fig:m-PU-WW_shift}, we show the central mass value and width as
a function of number of SD layers for different pileup levels.
One can see that as $N$ increases, the peak location improves while
the distribution becomes narrower, though the performance basically saturates at $N = 2$.

\begin{figure}[t]
  \centering
    \subfloat[]{%
    \includegraphics[trim={0 0 0 0.5cm},clip,width=0.46\textwidth]{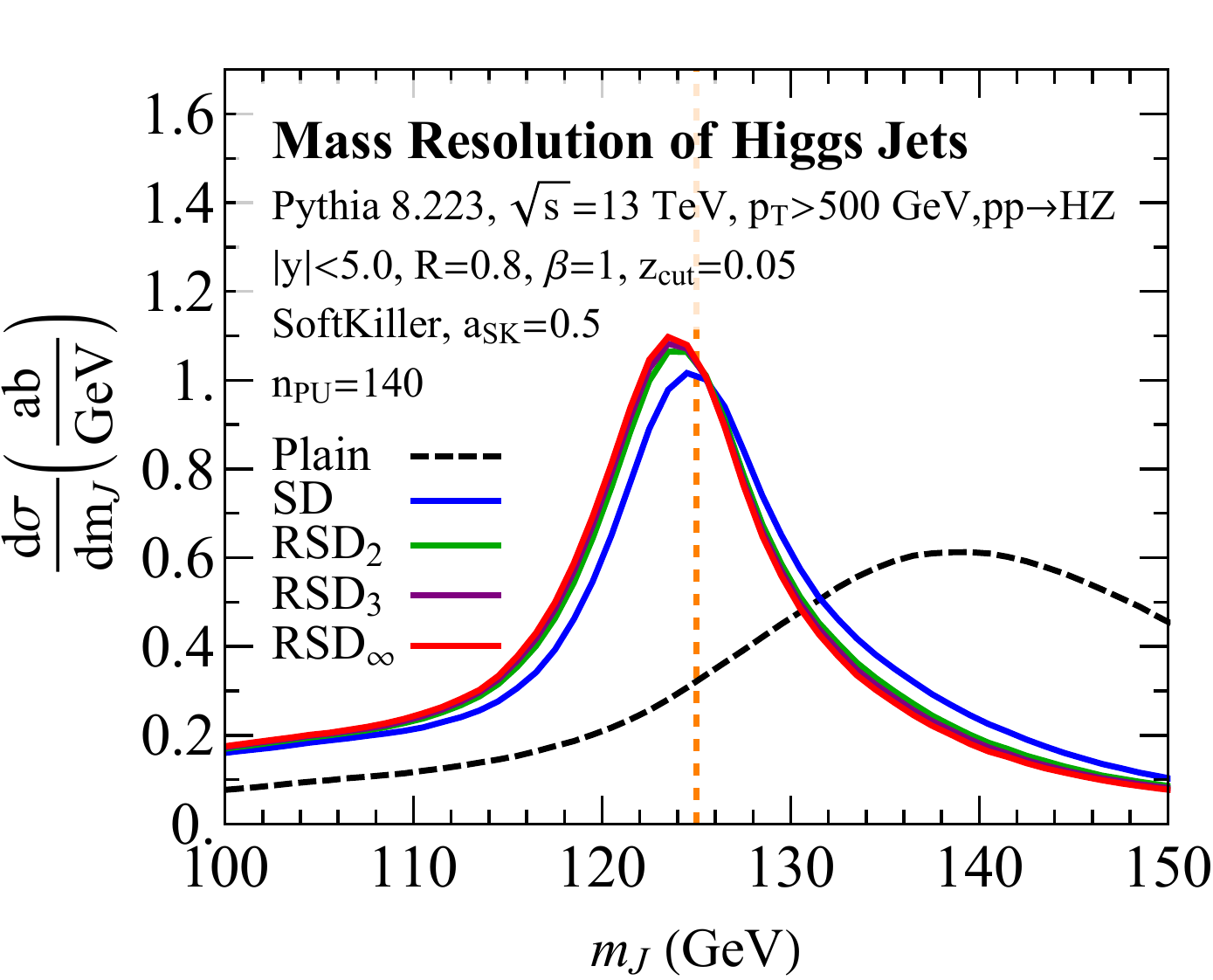}%
    \label{fig:m-PU-HZ_dist}}%
  \qquad
  \subfloat[]{\label{fig:m-PU-HZ_shift}%
  \begin{tabular}[b]{c}
    \includegraphics[trim={0 1.52cm 0 0},clip,width=0.4\textwidth]{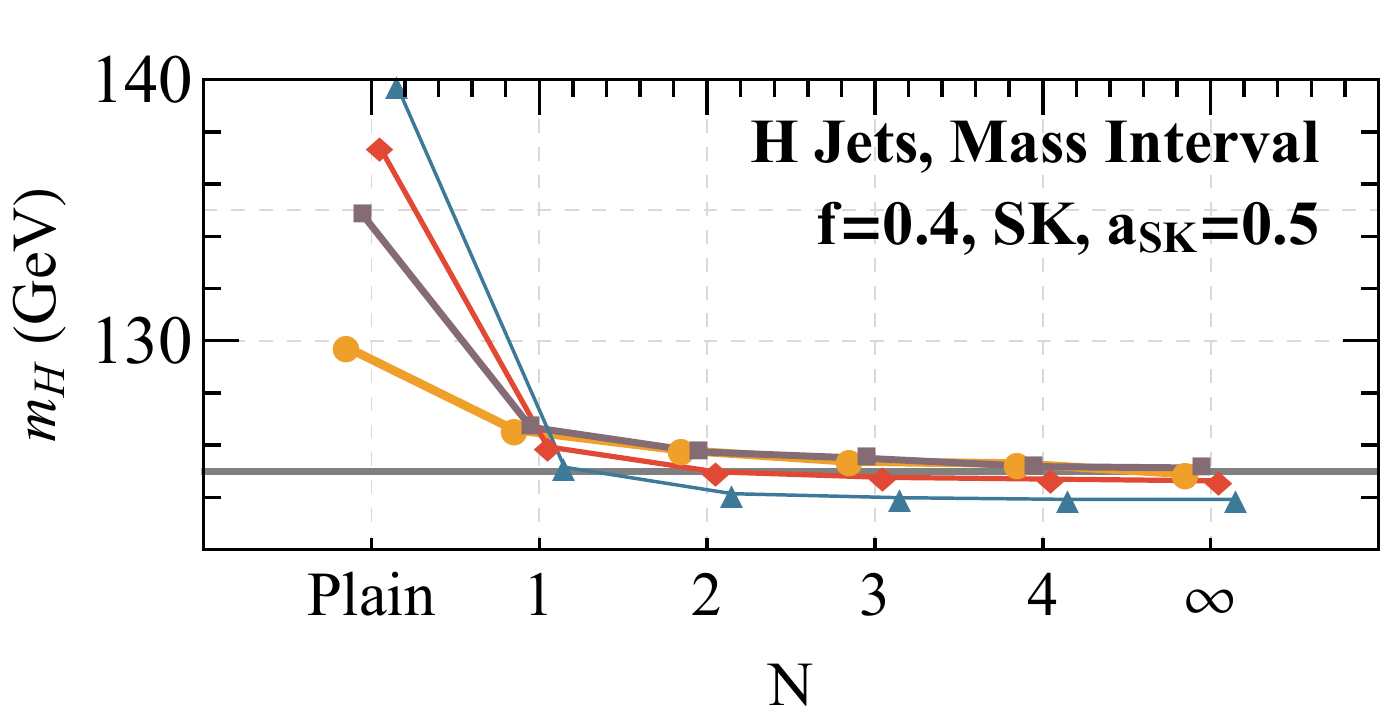} \\[2mm] % 
    \includegraphics[trim={0 0.2cm 0 0.5cm},width=0.39\textwidth]{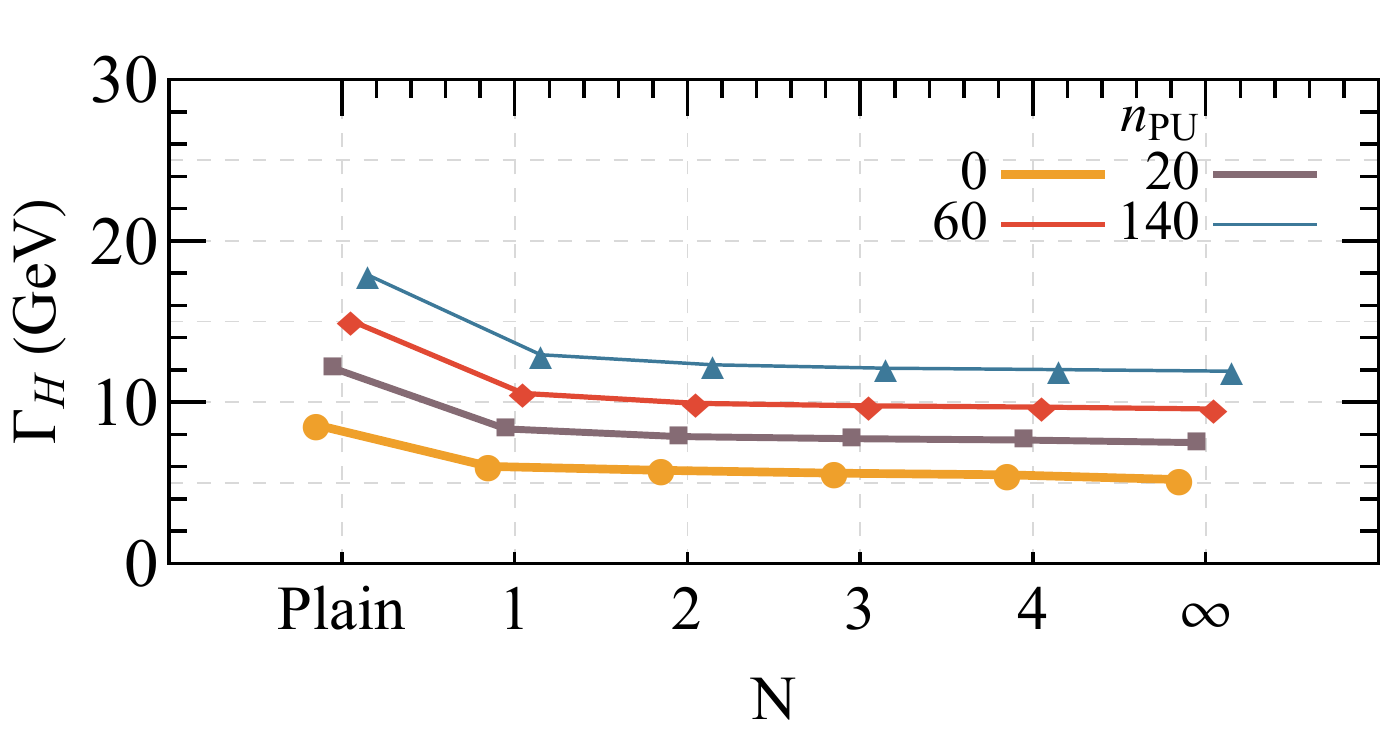}%
    \end{tabular}}
  \caption{Same as \Fig{fig:m-PU-top} but for the $H(\rightarrow4q)$ boson sample.}
  \label{fig:m-PU-HZ}
\end{figure}

In \Fig{fig:m-PU-HZ_dist} we show the mass distribution for $H\rightarrow 4q$ decays, again with 140 pileup vertices.
This is using the same event samples as in \Sec{sec:higgs_mass} and can be compared to \Fig{fig:H_mass_distribution_dist}.
Again, one can observe a small improvement in both the location of the central value and the width of the distribution as the number of
SD layers is increased.
This is shown more explicitly in \Fig{fig:m-PU-HZ_shift}, where we can see the narrowing of the
distribution and convergence of the peak for different pileup levels, with performance saturating around $N = 3$ or $N = 4$.

\subsection{Boosted top tagging with pileup}
\label{sec:toptag-pileup}

\begin{figure}
  \centering
    \includegraphics[width=0.5\textwidth]{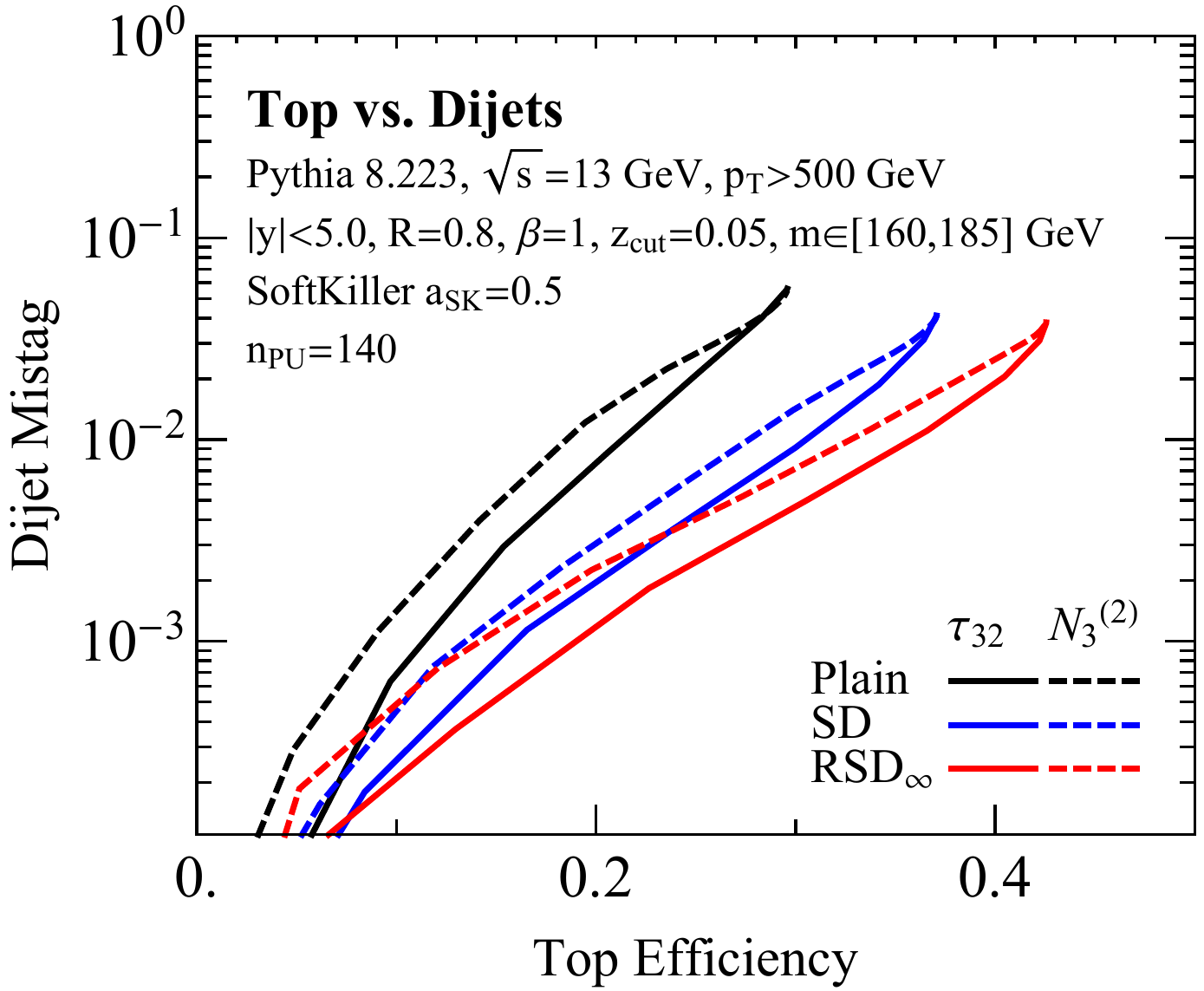}
    \caption{Same as~\Fig{fig:discrim-ROC}, but with 140 pileup
      vertices and SoftKiller with grid parameter $a_{\rm SK}=0.5$.}
  \label{fig:discrim-ROC-PU}
\end{figure}

Here, we repeat the top tagging study in \Sec{sec:boosted-toptag} in the presence of pileup, again using SoftKiller with a grid
parameter $a_{\rm SK}=0.5$.
In \Fig{fig:discrim-ROC-PU}, we show the same ROC curve as in \Fig{fig:discrim-ROC}, but with the addition of 140 pileup vertices.
As in the previous study without pileup, additional SD grooming layers improve the signal efficiency, with the observables
after RSD leading to much better discrimination between top and QCD jets.
Here, though, the improvement is much more substantial, since the gains in mass resolution in high-pileup condition is quite dramatic; see \Sec{sec:mass_PU}.
This allows for reasonable top tagging performance even with a narrow $m \in [160,185]\GeV$ top window, despite the large pileup multiplicity.

%======================================================================

\bibliographystyle{JHEP}
\bibliography{recursiveSoftdrop}

\end{document}